\documentclass[acmlarge, screen]{acmart}

\usepackage{tcolorbox}
\tcbuselibrary{skins, breakable, theorems}
\usepackage{times}
\usepackage{booktabs}
\usepackage{tabularx}
\usepackage{tabu}
\usepackage{multirow}
\usepackage{graphicx}
\usepackage{makecell}
\usepackage{subfigure}
\usepackage{amsmath}
\usepackage{bbding}
\usepackage{pifont}
\usepackage{wasysym}

\usepackage{amssymb}
\usepackage{lscape}
\usepackage{adjustbox}
\usepackage{framed}
\usepackage{color}
\usepackage{soul}
\def\BibTeX{{\rm B\kern-.05em{\sc i\kern-.025em b}\kern-.08emT\kern-.1667em\lower.7ex\hbox{E}\kern-.125emX}}

\acmJournal{CSUR}
\acmYear{2020}\acmVolume{1}\acmNumber{1}\acmArticle{1}\acmMonth{1}
\acmDOI{10.1145/1122445.1122456}

\begin{document}

%
% The "title" command has an optional parameter, allowing the author to define a "short title" to be used in page headers.
\title{A Survey on Deep Learning for Software Engineering}

%
% The "author" command and its associated commands are used to define the authors and their affiliations.
% Of note is the shared affiliation of the first two authors, and the "authornote" and "authornotemark" commands
% used to denote shared contribution to the research.
\author{Yanming Yang}
%\authornote{Both authors contributed equally to this research.}
\email{Yanming.Yang@monash.edu}
% \orcid{1234-5678-9012}
% \author{G.K.M. Tobin}
% \authornotemark[1]
% \email{webmaster@marysville-ohio.com}
\affiliation{%
  \institution{Faculty of Information Technology, Monash University}
%   \streetaddress{P.O. Box 1212}
  \city{Melbourne}
  \country{Australia}
%   \state{Ohio}
%   \postcode{43017-6221}
}

\author{Xin Xia}
\affiliation{%
  \institution{Faculty of Information Technology, Monash University}
%   \streetaddress{1 Th{\o}rv{\"a}ld Circle}
  \city{Melbourne}
  \country{Australia}}
\email{Xin.Xia@monash.edu}

\author{David Lo}
\affiliation{%
  \institution{School of Information Systems, Singapore Management University}
%   \city{Rocquencourt}
  \country{Singapore}}
  \email{davidlo@smu.edu.sg}

\author{John Grundy}
\affiliation{%
 \institution{Faculty of Information Technology, Monash University}
%  \streetaddress{Rono-Hills}
 \city{Melbourne}
%  \state{Arunachal Pradesh}
 \country{Australia}}
 \email{John.Grundy@monash.edu}

%
% By default, the full list of authors will be used in the page headers. Often, this list is too long, and will overlap
% other information printed in the page headers. This command allows the author to define a more concise list
% of authors' names for this purpose.

%\renewcommand{\shortauthors}{Trovato and Tobin, et al.}

%
% The abstract is a short summary of the work to be presented in the article.
\begin{abstract}
In 2006, Geoffrey Hinton proposed the concept of training ``Deep Neural Networks (DNNs)'' and an improved model training method to break the bottleneck of neural network development. More recently, the introduction of AlphaGo in 2016 demonstrated the powerful learning ability of deep learning and its enormous potential. Deep learning has been increasingly used to develop state-of-the-art software engineering (SE) research tools due to its ability to boost performance for various SE tasks. There are many factors, e.g., deep learning model selection, internal structure differences, and model optimization techniques, that may have an impact on the performance of DNNs applied in SE. Few works to date focus on summarizing, classifying, and analyzing the application of deep learning techniques in SE. To fill this gap, we performed a survey to analyse the relevant studies published since 2006. We first provide an example to illustrate how deep learning techniques are used in SE. We then summarize and classify different deep learning techniques used in SE. We analyzed key optimization technologies used in these deep learning models, and finally describe a range of key research topics using DNNs in SE. Based on our findings, we present a set of current challenges remaining to be investigated and outline a proposed research road map highlighting key opportunities for future work.
\end{abstract}

%
% The code below is generated by the tool at http://dl.acm.org/ccs.cfm.
% Please copy and paste the code instead of the example below.
%

% \begin{CCSXML}
% <ccs2012>
%  <concept>
%   <concept_id>10010520.10010553.10010562</concept_id>
%   <concept_desc>Computer systems organization~Embedded systems</concept_desc>
%   <concept_significance>500</concept_significance>
%  </concept>
%  <concept>
%   <concept_id>10010520.10010575.10010755</concept_id>
%   <concept_desc>Computer systems organization~Redundancy</concept_desc>
%   <concept_significance>300</concept_significance>
%  </concept>
%  <concept>
%   <concept_id>10010520.10010553.10010554</concept_id>
%   <concept_desc>Computer systems organization~Robotics</concept_desc>
%   <concept_significance>100</concept_significance>
%  </concept>
%  <concept>
%   <concept_id>10003033.10003083.10003095</concept_id>
%   <concept_desc>Networks~Network reliability</concept_desc>
%   <concept_significance>100</concept_significance>
%  </concept>
% </ccs2012>
% \end{CCSXML}

% \ccsdesc[500]{Computer systems organization~Embedded systems}
% \ccsdesc[300]{Computer systems organization~Redundancy}
% \ccsdesc{Computer systems organization~Robotics}
% \ccsdesc[100]{Networks~Network reliability}

%
% Keywords. The author(s) should pick words that accurately describe the work being
% presented. Separate the keywords with commas.

%
% A "teaser" image appears between the author and affiliation information and the body
% of the document, and typically spans the page.
%%\begin{teaserfigure}
%%  \includegraphics[width=\textwidth]{sampleteaser}
%%  \caption{Seattle Mariners at Spring Training, 2010.}
%%  \Description{Enjoying the baseball game from the third-base seats. Ichiro Suzuki preparing to bat.}
%%  \label{fig:teaser}
%%\end{teaserfigure}

%
% This command processes the author and affiliation and title information and builds
% the first part of the formatted document.
\maketitle

\section{Introduction}

In 1943, Warren Mcculloch and Walter Pitts first introduced the concept of the Artificial Neural Network (ANN) and proposed a mathematical model of an artificial neuron \cite{mcculloch1943logical}. This pioneered a new era of research on artificial intelligence (AI). In 2006, Hinton et al. \cite{hinton2006fast} proposed the concept of ``Deep Learning (DL)''. They believed that an ANN with multiple layers possessed extraordinary feature learning ability, which allows the feature data learned to represent the essence of the original data. In 2009, they proposed Deep Belief Networks(DBN) and an unsupervised greedy layer-wise pre-training algorithm \cite{mohamed2009deep}, showing great ability to solve complex problems. DL has since attracted attention of academics and industry practioners for many tasks. Development of Nvidia’s graphics processing units (GPUs) significantly reduced the computation time of DL-based algorithms. DL has now entered a period of great development. In 2012 Hinton's research group participated in an image recognition competition for the first time and won the championship in a landslide victory by training a CNN model called AlexNet on the ImageNet dataset. AlexNet outperformed the second best classifier (SVM) by a substantial margin. In March 2016, AlphaGo was developed by DeepMind, a subsidiary of Google, which defeated the world champion of Go by a big score. With continuous improvements in DL's network structures, training methods and hardware devices, DL has been widely used to solve a wide variety of research problems in various fields. 

%As tremendous achievements were made through incredibly dramatic efforts in the 2010s, DL has entered the period of high-speed development. In 2011, Glorot et al. \cite{glorot2011deep} proposed the ReLU activation function, which can effectively suppress the vanishing gradient problem.

Driven by the success of DL techniques in image recognition and data mining, industrial practitioners and academic researchers have shown great enthusiasm for exploring and applying DL algorithms in diverse software engineering (SE) tasks, including requirements, software design and modeling, software implementation, testing and debugging, and maintenance. In requirements engineering, various DL algorithms have been employed to extract key features for requirement analysis, and automatically identify actors and actions (i.e., user cases) in natural language-based requirement descriptions \cite{wiegers2013software, al2018use, pudlitz2019extraction}. In software design and modeling, DL has been leveraged for design pattern detection \cite{thaller2019feature}, UI design search \cite{chen2020wireframe}, and software design mining \cite{mahadi2020cross}. During software implementation, researchers and developers have used DL for source code generation \cite{gao2019neural}, source code modeling \cite{hussain2020codegru}, software effort/cost estimation \cite{bisi2016software}, etc. In software testing and debugging, various DL algorithms have been designed for detecting and fixing defects and bugs existed in software products, e.g., defect prediction \cite{wang2016automatically}, bug localization \cite{li2020deep}, vulnerability prediction \cite{han2017learning}. It has been used for a varietu of software testing applications, such as test case generation \cite{liu2017automatic}, and automatic testing \cite{zheng2019wuji}. Researchers have applied DL to SE tasks to facilitate software maintenance and evolution, such as code clone detection \cite{nafi2019clcdsa}, feature envy detection \cite{liu2018deep}, code change recommendation \cite{siow2020core}, user review classification \cite{genc2017systematic}, etc.

%As DL has been used extensively for various SE tasks due to its benefits (e.g., less feature engineering effort, higher accuracy), \hl{some controversial studies have noticed some challenges cannot be ignored when training a DL-based model (e.g., longer training time/testing time, overfitting/underfitting problem, and the explanation of DL models), and the performance of DL algorithms are unable to outperform typical machine learning (ML) algorithms in certain SE tasks, which reminders us not to abuse the DL technique.} For example, Fu and Menzies \cite{fu2017easy} proposed a simple optimizer called DE to fine-tune SVM, achieving similar (or even better) performance as a DL solution, while reducing the training time substantially. Liu et al. \cite{liu2020replicability} performed a literature review to investigate the replicability and reproducibility of DL techniques. They observed that only 10.8\% of the studies discussed DL' replicability and/or reproducibility, and more than 74.2\% of the studies did not provide any replication package. Majumder et al. \cite{menzies2018500+} proposed a new approach to generate classifiers by fine-tuning, and the generated classifiers are nearly as good as DL models and easier to reproduce with faster running time and fewer CPU resources.

However, there is a lack of a comprehensive survey of deep learning usage to date in SE. %For helping practitioners and researchers get an overview of DL and providing them useful strategies to adopt or develop practical DL algorithms correctly, we need to draw a clear road map of the research conducted in the intersection of the two areas: DL \& SE.
This study performs a detailed survey to review, summarize, classify, and analyze relevant papers in the field of SE that apply DL models. We collected, reviewed, and analyzed 142 papers published in 20 major SE conferences and journals since ``deep learning'' was introduced in 2006. We then analyzed the development trends of DL in SE, classified various DL techniques used in diverse SE tasks, analyzed DL's construction process, and summarized the research topics tackled by relevant papers. This study makes the following contributions:

%Based on our findings, we highlight a set of current challenges that still need to be addressed and provide some key potential opportunities for further research.

\begin{enumerate}
    \item We conducted a detailed analysis on 142 relevant studies that used DL techniques in terms of publication trend, distribution of publication venues, and types of contributions. We analyzed an example in detail to describe the basic framework and the usage of DL techniques in SE.
    
    %\item We provide an overview of DL technique usage in SE, in terms of the publication trend, publication venues, and the main contribution of each primary study.
    
   \item We provide a classification of DL models used in SE based on their architectures and  an analysis of DL technique selection strategy.
    
    \item We performed a comprehensive analysis on the key factors that impact the performance of DL models in SE, including dataset, model optimization, and model evaluation. 
    
    \item We provide a description of each primary study according to six different SE activities and conducted an analysis on these studies based on their task types. These include regression, classification, ranking, and generation tasks. %We also listed many topics where DL techniques have been widely used.

    \item We discuss distinct technical challenges of using DL in software engineering and outline key future avenues for research on using DL in software engineering.
\end{enumerate}

Section 2 introduces the workflow of a DL model through an example. Section 3 presents our Systematic Literature Review methodology. Section 4 investigates the distribution and evolution of DL studies for SE tasks, and Section 5 gives an overall analysis on various DL techniques used in primary studies, including classifying 30 DNN models based on their architectures and summarizing the model selection strategies adopted by studies. Section 6 analyzes a set of key techniques from four perspectives -- datasets, model optimization, model evaluation, and the accessibility of source code. Section 7 lists research topics involved in primary studies and makes a briefly description of each work. Section 8 presents limitations of this study and its main threats to validity. Section 9 discusses the challenges that still need to be solved in future work and outlines a clear research road-map of research opportunities. Section 10 concludes this paper.

\section{Deep learning}

%In this section, to get a better sense of DL, we first briefly introduced the basic structure of a DL model and then analyzed a specific work to outline the workflow of the DL techniques used in SE.

\subsection{Basic Structure of DL}

Most learning algorithms are shallow models with one or two non-linear feature representation layers, such as GMM, HMM, SVM, and MLP. The limitation of such a shallow model is the lack of  ability to express complex functions. Their generalization ability is restricted for the complexity of problems, resulting in decreased learning ability.

\begin{figure} %\vspace{-0.3cm}
    \centering
    \includegraphics[width=0.6\textwidth]{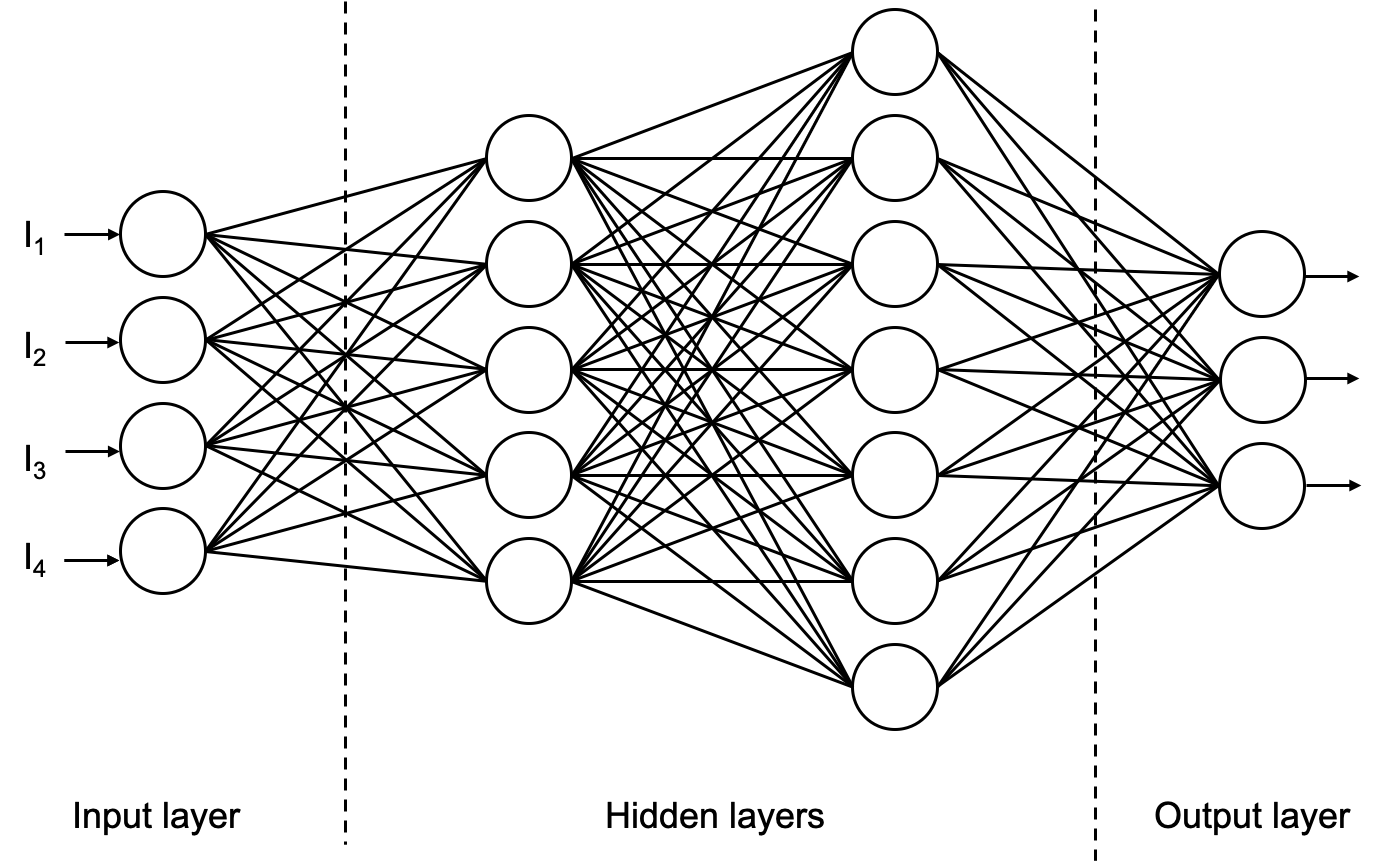}
   \vspace{-0.3cm} \caption{The basic structure of a deep learning model.}\vspace{-0.5cm}
    \label{fig:DL_model}
\end{figure}

Deep learning allows computational models composed of multiple layers to learn data representations with multiple higher levels of abstraction \cite{lecun2015deep}. This builds a neural network that simulates the human brain for analysis and learning. Similar to traditional ML, DL is suitable for various types of problems, such as regression, clustering, classification, ranking, and generation problems. We present the basic structure of a DNN in Fig. \ref{fig:DL_model}.

Based on the position and function of different layers, layers in a DNN can be classified into three categories, i.e., the input layer, the hidden layer, and the output layer. Generally, the first layer denotes the input layer, where the preprocessed data can be fed into DNNs; the last layer denotes the output layer, from which the results of a model can be achieved, e.g., classification results, regression results, generation results, etc. The middle layers between the input layer and the output layer are hidden layers. DNNs usually contain multiple hidden layers for enhancing the expressive ability of DNNs and learning high-level feature representation. Besides, the way to connect between different layers may vary, and as shown in Fig. \ref{fig:DL_model}, adjacent layers are full-connected (aka., full-connected layer), meaning that any neuron in the $i_{th}$ layer are connected to any neuron in the $i+1_{th}$ layer. In some DNNs with complex structures, for tackling different SE issues, not only the fully connected layer can be used as a hidden layer, but also a layer composed of other types of neurons can also be used as the hidden layer of a DNN, such as convolution layer, pooling layer, LSTM layers, etc..

Currently, diverse DNNs and learning methods are used for SE tasks, such as Feedback Neural Network (FNN), Convolutional Neural Networks (CNN), Recurrent Neural Networks (RNN), AutoEncoders, Generative Adversarial Networks (GAN), and Deep Reinforcement Learning \cite{goodfellow2016deep}.

\subsection{Workflow when Using DL in SE}

We present an example of using DL for a representative SE task. A novel sequence-to-sequence neural network is used to automatically perform comment updates with code changes \cite{liu2020Automating}.
Code comments are a vital source of software documentation, helping developers to have a better understanding of source code and are beneficial to the communication between developers. However, developers sometimes neglect to update comments after changing source code, leading to obsolete or inconsistent comments. It is necessary to perform Just-In-Time (JIT) code comments updating, which aims to reduce and even avoid bad comments. To achieve this, a new approach called CUP builds a novel neural seq2seq model. Using DL techniques for such an SE task can be broken into six steps: (1) data collection, (2) data processing, (3) input construction, (4) model training, (5) model optimization, (6) model testing, and (6) model application.

%\begin{enumerate}
     \textbf{Data collection:} Collecting data is a key step when building and training a DL model. Different SE tasks need to process and analyze different data types, such as requirements documents for requirements analysis, bug reports for bug report summarization, source code for code clone detection or code search, etc. In this example, for updating JIT code comments, the commonly available data are method-level code change datasets with comments. Each qualified instance contains old code snippets, new code snippets, old code comments, and new code comments.

    \textbf{Data processing:} Processing raw data involves a number of steps, including data filtering, data splitting, data cleaning, data augmentation, and data segmentation for eliminating noise in data. For JIT code comments updating, some instances are removed with unqualified comments, and no differences or empty between old and new comments. Instances containing abstract methods are deleted to reduce method mismatching.

    \textbf{Input construction:} Since most DL models have strict requirements on the input-form, such as the input requiring to be numeric and fixed size, it is necessary to transform SE data into multi-dimensional vectors in order to use DL models. In this JIT code comment updating example, code comments and code changes can be viewed as text-based data, processed by using a token-based method. Code changes and code comments are converted into sequences so that they can be fed into their seq2seq model through data flattening and tokenization. After tokenization, old comments are converted into a token sequence. While to better represent code changes,  each change is aligned into two token sequences and construct a triple $<t_i, t_i^{'}, a_i>$ as an edit sequence to respectively record old source code, new source code, and an edit action, i.e., insert, delete, equal or replace.

    \textbf{Model training:} DL users need to select suitable DL techniques and different datasets, construct the structure of a model, and decide model configuration, e.g., the number of layers and neural units of each layers. In our example, a seq2seq DL model was built by training an encoder-decoder LSTM, since it is good at process nature language text and token-based source code. In this model, the edit sequence of a code change and a token sequence of its old comment were fed into the input layer. To capture the relationship between the code change and the old comment, the encoder was composed of 4 layers: an embedding layer, a contextual embed layer, a co-attention layer, and a modeling layer, where each layer had its role. The decoder included 2 layers: a LSTM layer and a dense layer. The output of decoder was a new comment based on the corresponding captured relationship.

    \textbf{Model optimization:} After model design and construction, the designed model will be trained with the training set for achieving an effective DL model. Whether a model can work depends on thousands of parameters (aka., weights), connecting neural units adjacent layers. Hence, model training is to fine-tune these weights to minimize the loss of the model. For the seq2seq model in the example, the weights in the seq2seq neural network are trained by minimizing the difference between the real new comment and the generated new comment in a supervised way.

    \textbf{Model testing:} Generally, a training set is usually divided into two subsets of unequal sizes. The big subset is used for training the DL model, while the small one will be used for validating and testing the performance of the model when meeting new data. In this example, 20\% of samples in the training set are put into the validation and test set to ensure the effectiveness of CUP.

    \textbf{Model application:} Finally, the trained DL model can be applied to tackle practical SE tasks. In this example, the trained model leverages two distinct encoders and a co-attention layer to learn the relationships between the code change and the old comment. The LSTM-based decoder is used to generate new comments whose tokens are copied from both the new code and the old comments.
%\end{enumerate}

% \begin{tcolorbox}[breakable,colback=gray!20,colframe=gray!35!black,title=Summary]
%   To sum up, the DL technique has been one of the most prevalent options of model selections for various SE tasks as the evolution of DL. Usually, a DNN consists of three types of layers, i.e., an input layer, multiple hidden layers, and an output layer. There are many ways to connect between adjacent layers and various types of NN layers can be used as the hidden layer, which increases the number of DNN types.  For introducing the workflow of the DNN in SE, We also give a detailed example that leverages multiple DNN techniques to automatically updating old code comments based on code changes.
% \end{tcolorbox}

\section{Methodology}

We performed a systematic literature review (SLR) following  Kitchenham and Charters \cite{keele2007guidelines} and Petersen et al. \cite{petersen2015guidelines}. In this section, we present details of our SLR methodology.

\subsection{Research Questions}

We want to analyse the history of using DL models in SE by summarizing and analyzing the relevant studies, and providing the guidelines on how to select and apply the DL techniques. To achieve this, we wanted to answer the following research questions:

\begin{enumerate}
    \item \textbf{RQ1: What are the trends in the primary studies on the use of DL in SE?}
    \item \textbf{RQ2: What DL techniques have been applied to support SE tasks?}
    \item \textbf{RQ3: What key factors contribute to difficulties in training DNNs for SE tasks? }
    \item \textbf{RQ4: What types of SE tasks and which SE phases have been facilitated by DL-based approaches?}
\end{enumerate}

RQ1 analyzes the distribution of relevant publications that used DL in their studies since 2006 to give an overview of the trend of DL in SE. RQ2 provides a classification of different DL techniques supporting SE tasks and analyze their popularity based on their frequency of use in SE. RQ3 explores key technologies and factors that may affect the efficiency of the DNN training phase. RQ4 investigates what types of SE tasks and which SE phases have been facilitated by DNNs.

\subsection{Literature search and selection}

To collect DL related papers in SE, we identified a search string including several DL related terms frequently appeared in SE papers that make use of DL. We then refined the search string by checking the title and abstract of a small number of relevant papers. After that, we used logical ORs to combine these terms, and the search string is:

% \textit{
% \footnotesize
% ("deep" OR "neural" OR "Intelligence" OR "reinforcement")
% }

("deep" OR "neural" OR "Intelligence" OR "reinforcement")

\begin{table*}[htbp]
\centering
\footnotesize
\caption{Publication venues for manual search}
\label{tab:venues}
 \vspace{-0.3cm}\begin{tabular}{llp{10cm}}
  \toprule
 \textbf{No.} & \textbf{Acronym} & \textbf{Full name}\\
  \midrule
 %conference
 1. & ICSE & ACM/IEEE International Conference on Software Engineering  \\
 2. & ASE  & IEEE/ACM International Conference Automated Software Engineering \\
 3. & ESEC/FSE  & ACM SIGSOFT Symposium on the Foundation of Software Engineering/European Software Engineering Conference \\
 4. & ICSME & IEEE International Conference on Software Maintenance and Evolution \\
 5. & ICPC & IEEE International Conference on Program Comprehension \\
 6. & ESEM & International Symposium on Empirical Software Engineering and Measurement \\
 7. & RE & IEEE International Conference on Requirements Engineering \\
 8. & MSR & IEEE Working Conference on Mining Software Repositories \\
 9. & ISSTA & International Symposium on Testing and Analysis Working Conference on Mining Software Repositories \\
 10. & SANER & IEEE International Conference on Software Analysis, Evolution and Reengineering \\
 11. & ICST & IEEE International Conference on Software Testing, Verification and Validation \\
 12. & ISSRE & IEEE International Symposium on Software Reliability  Engineering \\
\midrule
 %journal
 13. & TSE & IEEE Transactions on Software Engineering \\
 14. & TOSEM & ACM Transactions on Software Engineering and Methodology \\
 15. & ESE & Empirical Software Engineering \\
 16. & JSS & Journal of Systems and Software \\
 17. & IST & Information and Software Systems \\
 18. & ASEJ & Automated Software Engineering \\
 19. & IETS & IET Software \\
 20. & STVR & Software Testing, Verification and Reliability \\
 21. & JSEP & Journal of Software: Evolution and Process \\
 22. & SQJ & Software Quality Journal \\
  \bottomrule
 \end{tabular}\vspace{-0.5cm}
\end{table*}

We specified the range the papers are published later: 2006-July 2020. Following previous studies \cite{hosseini2017systematic, huang2019survey, liu2020replicability}, we selected 22 widely read SE journals (10) and conferences (12) listed in Table \ref{tab:venues} to conduct a comprehensive literature review. We run the search string on three databases (i.e., ACM digital library \footnote{https://dl.acm.org}, IEEE Explore \footnote{https://ieeexplore.ieee.org}, and Web of Science \footnote{http://apps.webofknowledge.com}) looking for publications in the 22 publication venues whose meta data (including title, abstract and keywords) satisfies the search string. Our search returns 655 relevant papers. After discarding duplicate papers, we applied some inclusion/exclusion criteria (presented in Section 3.3) by reading their title, abstract and keywords, and narrow the candidate set to 146 studies. After reading these 146 studies in full to ensure their relevance, we retained 142 studies.

% In order to collect the relevant papers as fully as possible, we used the designed search string to perform an automatic search for DL related papers published since 2006 (i.e., the time to introduce the concept of ``Deep Learning'').

%Based on the search result, we downloaded relevant papers from three digital library portals, including ACM digital library \footnote{https://dl.acm.org}, IEEE Explore \footnote{https://ieeexplore.ieee.org}, and Web of Science \footnote{http://apps.webofknowledge.com}, and obtained 655 candidate studies.

\subsection{Inclusion and Exclusion Criteria}

After retrieving studies that match our search string, it is necessary to filter unqualified studies, such as studies with insufficient contents or missing information. To achieve this, we applied our inclusion and exclusion criteria to determine the quality of candidate studies for ensuring that every study we kept implemented and evaluated a full DL approaches to tackle SE tasks.

The following inclusion and exclusion criteria are used:

%\ding{52} \ The paper must be published between 2006-2020.

\ding{52} \ The paper must be written in English.

%\ding{52} \ The paper must be a peer reviewed full research paper published in the predefined publication venues.

\ding{52} \ The paper must adopt DL techniques to address SE problems.

\ding{52} \ The length of paper must not be less than 6 pages.

\ding{56} \ Books, keynote records, non-published manuscripts, and grey literature are dropped.

\ding{56} \ If a conference paper has an extended journal version, the conference version is excluded.

\subsection{Data Extraction and Collection}

After removing the irrelevant and duplicated papers, we extracted and recorded the essential data and performed overall analysis for answering our four RQs. Table \ref{tab:data} described the detailed information being extracted and collected from 142 primary studies, where the column $'Extracted Data Items'$ lists the related data items that would be extracted from each primary study, and the column $‘RQ’$ denotes the related research questions to be answered by the extracted data items on the right. To avoid making mistakes in data collection, two researchers extracted these data items from primary studies together and then another researcher double checked the results to make sure of the correctness of the extracted data.

%\hl{where the column $‘RQ’$ denotes the related research questions to be answered by the extracted data items on the right. In order to prevent data loss and avoid mistakes as can as possible, data collection was performed by two authors and then the results were verified by other researchers who are not co-authors of this paper.}

\begin{table}[htbp]
\centering
\footnotesize
\caption{Data Collection for Research Questions}
\label{tab:data}
\vspace{-0.3cm} \begin{tabular}{p{1cm}p{11cm}}
  \toprule
 \textbf{RQs} & \textbf{Extracted data items} \\
  \midrule
 RQ1 & Basic information of each primary study (i.e., title, publication year, authors, publication venue) \\
 RQ1 & The type of main contribution in each study (e.g., empirical study, case study, survey, or algorithm)\\

 RQ2 & DL techniques used in each study \\
 RQ2 & Whether and how the authors describe the rationale behind techniques selection \\

 RQ3 & Dataset source (e.g., industry data, open source data, or collected data) \\
 RQ3 & Data type (e.g., source code, nature language text, and pictures) \\
 RQ3 & {The process that datasets are transformed into input sets suitable for DNNs} \\
%  RQ3 & Whether, what and how data processing techniques are used \\
% RQ3 & Which a deep learning platform is used to implement DNNs. (e.g., tensorflow, keras, and pycharm) \\
 RQ3 & Whether and what optimization techniques are used \\
% RQ3 & Whether the study evaluate the runtime of the DL model \\
 RQ3 & What measures are used to evaluate the DL model \\
 RQ3 & Presence / absence of replication package \\

 RQ4 & The practical problem that a SE task tries to solve\\
 RQ4 & The SE activity in which each SE task belongs\\
 RQ4 & {The approach used for each SE task} (e.g., regression, classification, ranking, and generation) \\
  \bottomrule
 \end{tabular}\vspace{-0.3cm}
\end{table}

\section{RQ1: What are the trends in the primary studies on use of DL in SE?}

We analyzed the basic information of primary studies to comprehend the trend of DL techniques used in SE in terms of the publication date, publication venues, and main contribution types of primary studies.

\subsection{Publication trends of DL techniques for SE}

\begin{figure}%[htbp]
\centering
\subfigure[Number of publications per year.]{
\centering
\begin{minipage}[t]{0.42\textwidth}
\includegraphics[width=1\textwidth]{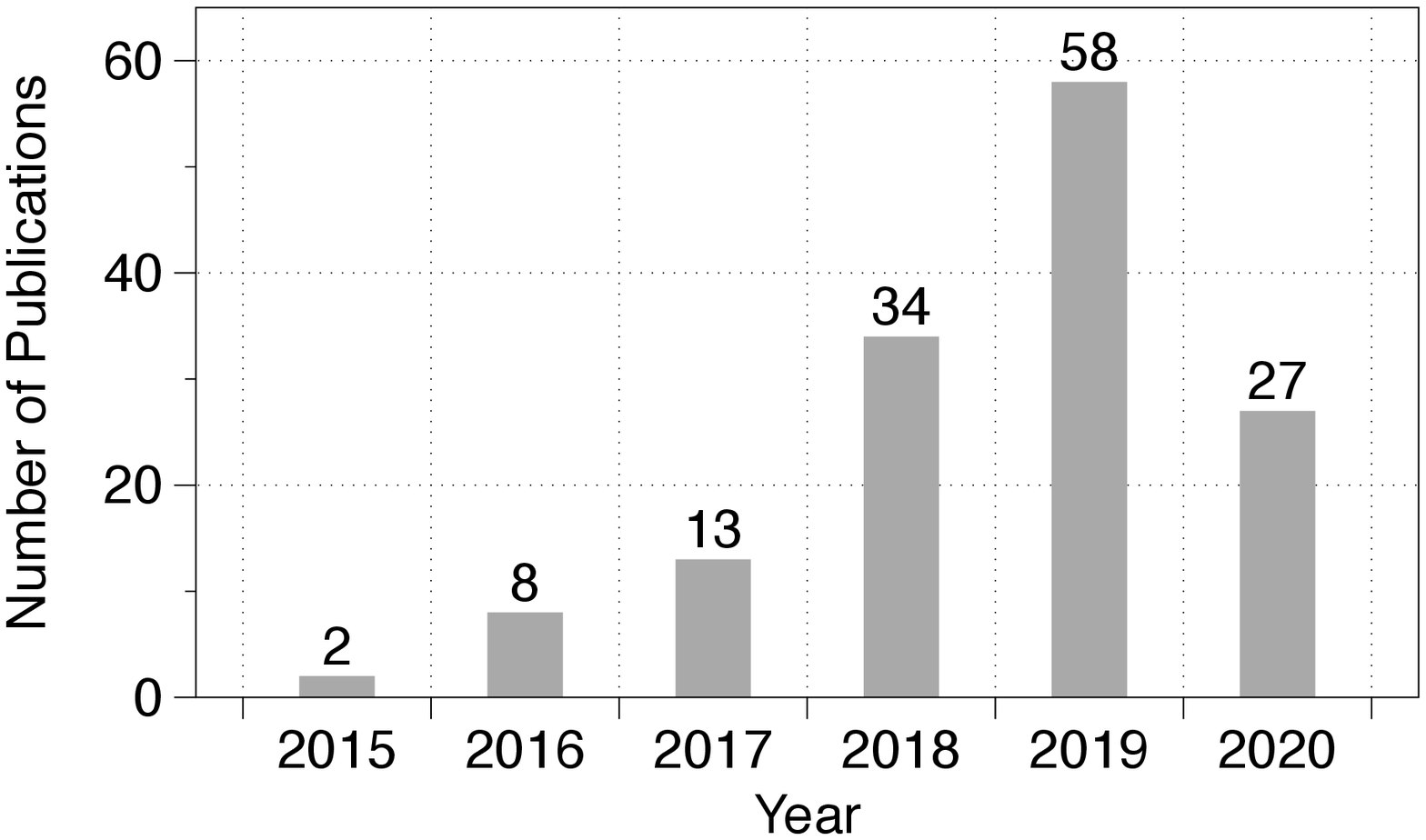}
\end{minipage}
}
%\hfill
\subfigure[Cumulative number of publications per year.]{
%\centering
\begin{minipage}[t]{0.42\textwidth}
\includegraphics[width=1\textwidth]{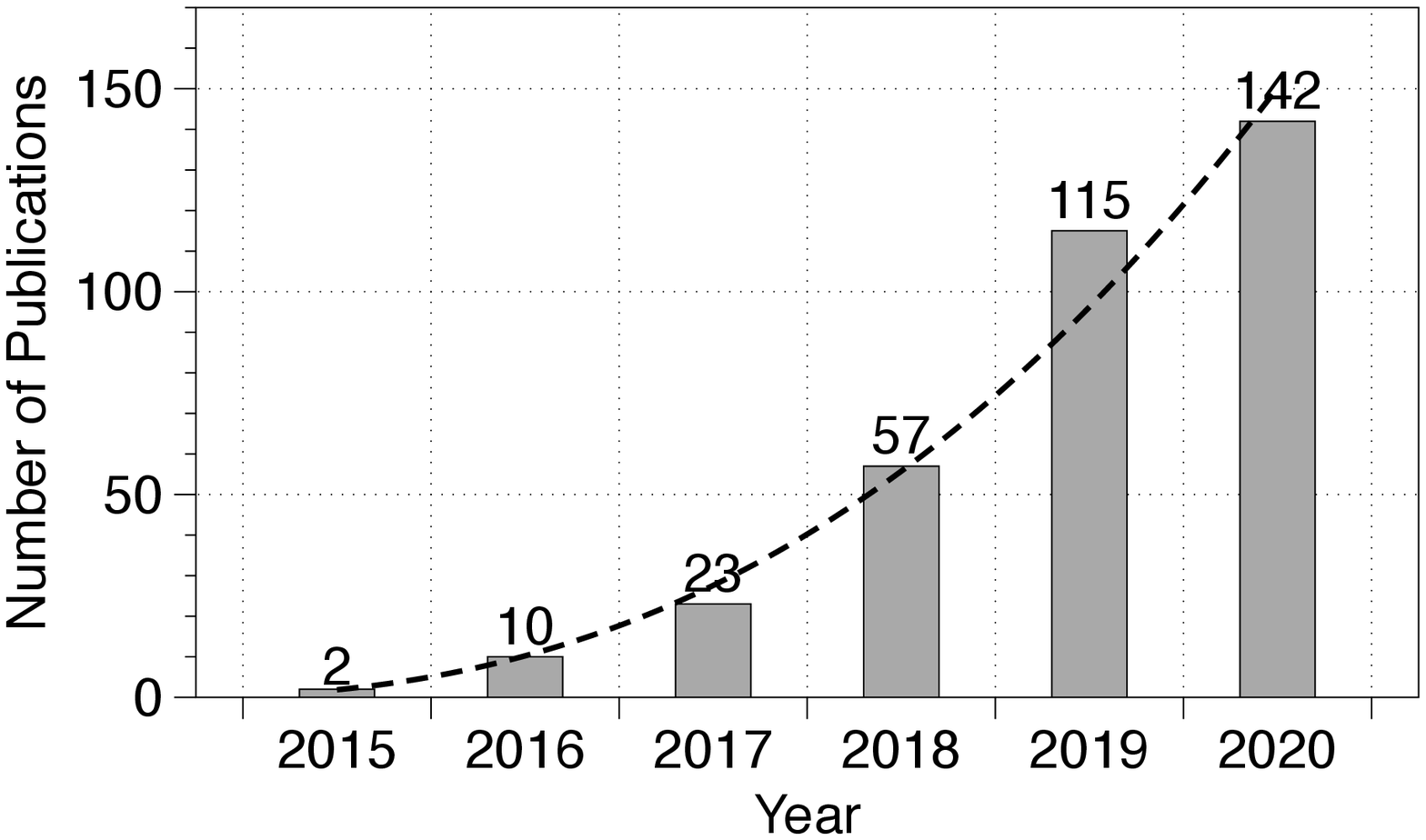}
\end{minipage}
}
\vspace{-0.3cm}\caption{Publication trends of DL-based primary studies in SE.}\vspace{-0.5cm}
\label{fig:publication_trend}
\end{figure}

We analyzed the publication trends of DL-based primary studies published between 2006 and the middle of 2020. Although the concept of ``Deep Learning'' has been proposed in 2006 and DL techniques had been widely used in many other fields in 2009, we did not find any studies using DL to address SE tasks before 2015. Fig. \ref{fig:publication_trend}(a) shows the number of relevant studies published in predefined publication venues since the middle of 2020. It can be observed that the number of publications from 2015 to 2019 shows a significant increase, with the number reaching 58 papers in 2019. In data collection, we only collect papers whose initial publication date is on July 2020 or earlier; thus, the number of relevant studies in 2020 cannot reveal the overall trend of DL in 2020. However, extrapolating from the number of primary studies in previous years, we can estimate that there may be over 65 relevant publications using various DL techniques to solve SE issues by the end of 2020.

We also performed an analysis of the cumulative number of  publications as shown in Fig. \ref{fig:publication_trend}(b). We fit the cumulative number of publications as a power function, showing the publication trend in the last five years. We can notice that the slope of the curve fitting the distribution increases substantially between 2015 and 2019, and the coefficient of determination ($R^{2}$) attains the peak value (0.99447), which indicates that the number of relevant studies using DL in SE intends to experience a strong rise in the future. Therefore, after analyzing Fig. \ref{fig:publication_trend}, it can be foreseen that using DL techniques to address various SE tasks has become a prevalent trend since 2015, and huge numbers of studies will adopt DL to address further challenges of SE.

\subsection{Distribution of publication venues}

We reviewed 142 studies published in various publication venues, including 12 conference proceedings and symposiums as well as 10 journals, which covers most research areas in SE. Table \ref{tab:venue} lists the number of relevant papers published in each publication venue. 69\% of publications appeared in conferences and symposiums, while only 31\% of journal papers leveraged DL techniques for SE tasks. Among all conference papers, only 4 different conferences include over 10 studies using DL in SE in the last five years, i.e., SANER, ASE, ICSE, and MSR. Compared with other conference proceedings, SANER is the most popular one containing the highest number of primary study papers (22), followed by ASE (16). There are 13 and 6 relevant papers published in ICSE and FSE, respectively. Meanwhile, in all journals, TSE and IST include the highest number of relevant papers (11). Ten studies related to DL techniques were published in JSS, and 5 were published in TOSEM. Almost half of the publication venues only published not more than 5 relevant papers.

\begin{table} 
\centering
\footnotesize
\caption{ Publication Venues with DL-based Studies.}
\label{tab:venue}
\vspace{-0.3cm}\begin{tabular}{cccc}
\toprule
\textbf{Conference venue} & \textbf{\#Studies} & \textbf{Journal venue} & \textbf{\#Studies}\\
   \midrule
SANER  & 22   & TSE    &  11  \\
ASE    & 16   & IST    &  11  \\
ICSE   & 13   & JSS    &  10  \\
MSR    & 10   & TOSEM  &  5   \\
ICSME  &  8  & ESE    &  2   \\
ISSTA  &  7   & ASEJ   &  2   \\
FSE    &  6   & IETS   &  1   \\
ICST   &  5   & STVR   &  1   \\
ICPC   &  4    & SQJ    &  1   \\
RE     &  3 & &  \\
ISSRE  &  3 & &  \\
ESEM  &  1  & &\\

  \bottomrule
\end{tabular}\vspace{-0.5cm}
\end{table}

\begin{figure} 
\centering
\subfigure[Number of primary studies published in various conference proceedings.]{
\centering
\begin{minipage}[t]{0.48\textwidth}
\includegraphics[width=1\textwidth]{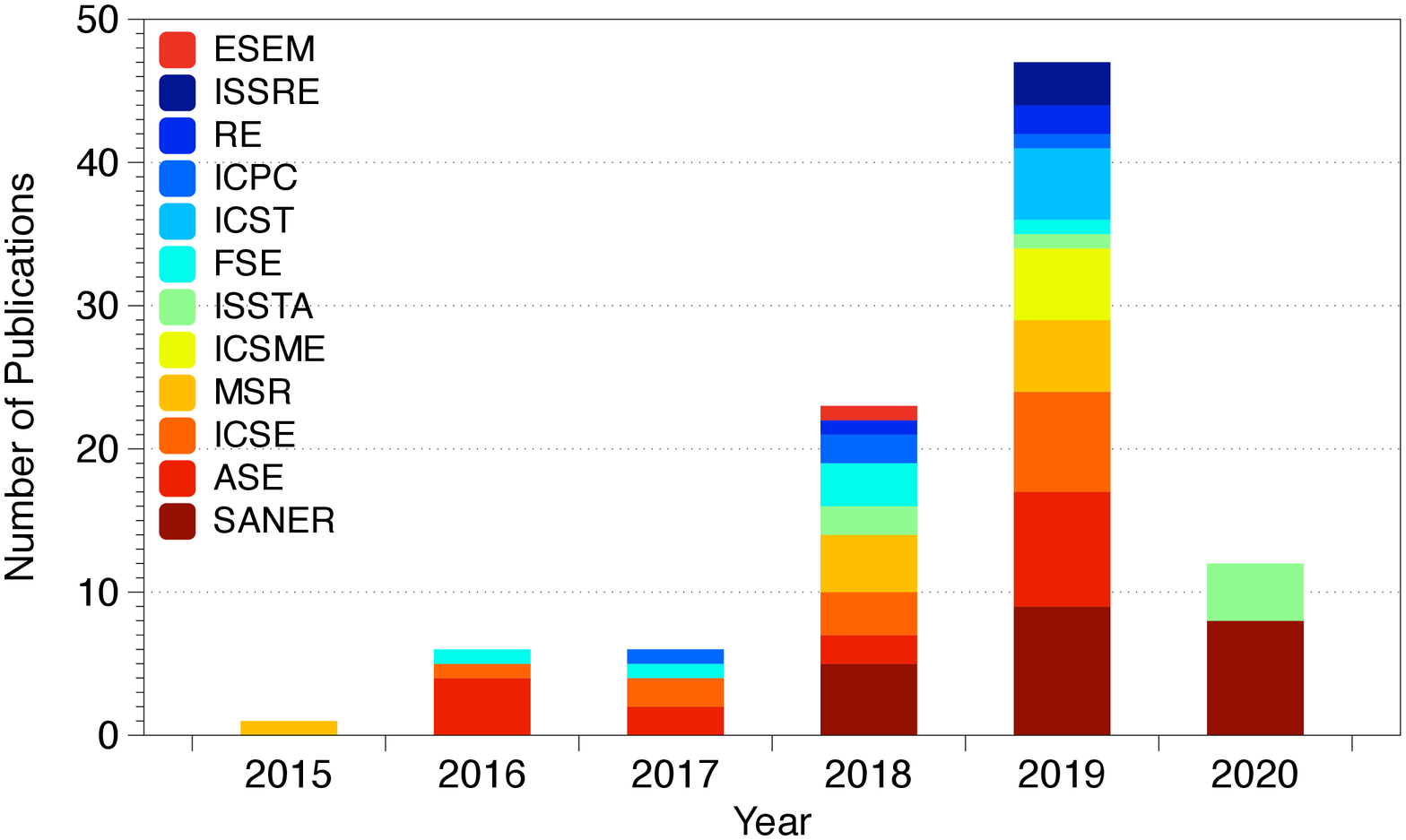}
\end{minipage}
}
%\hfill
\subfigure[Number of primary studies published in various journals.]{
%\centering
\begin{minipage}[t]{0.48\textwidth}
\includegraphics[width=1\textwidth]{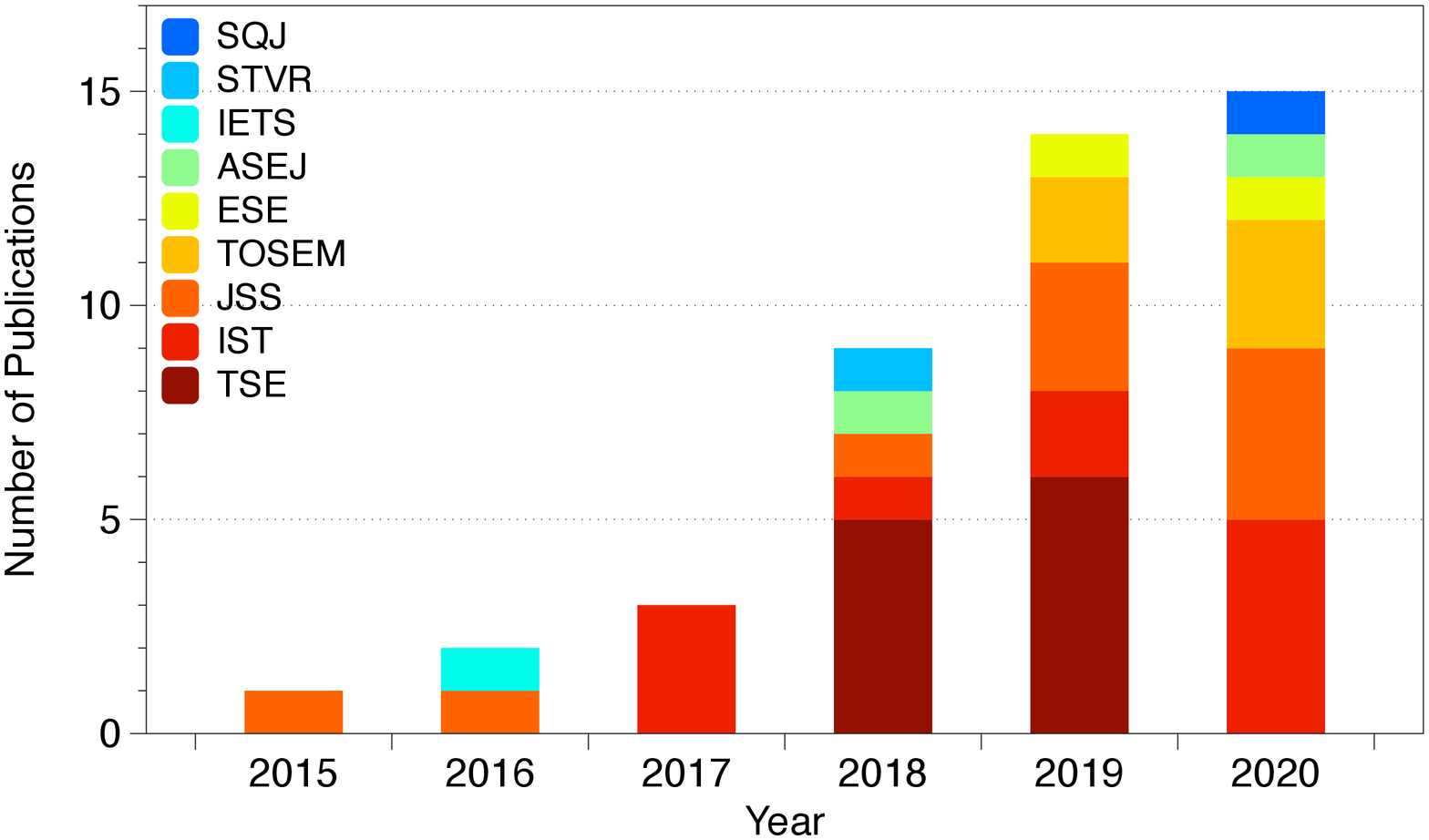}
\end{minipage}
}
\vspace{-0.3cm}\caption{Distribution of different publication venues.}
\label{fig:conference_and_journal}\vspace{-0.6cm}
\end{figure}

We also checked the distribution of primary studies published in conferences and journals between 2015 and 2020, shown in Fig. \ref{fig:conference_and_journal}. Fig \ref{fig:conference_and_journal}(a) illustrates that the publication trend of various conference proceedings and symposiums has a noticeable increase from 2015 to 2019. 70.4\% of conference papers were published in 2018 and 2019, while only a few different conferences or symposium venues included relevant papers between 2015 and 2017, which demonstrates a booming trend in the last few years.% We can observe that not a lot of primary study papers were published in 2020 and they were both included in SANER and ISSTA. The reason for this condition is that accepted papers of most conference venues in 2020 were not available online when collecting publications.

Fig. \ref{fig:conference_and_journal}(b) shows the number of primary study papers published in different journal venues. It can be seen that there is an increasing trend in the last five years, especially between 2018 and 2020. Furthermore, the relevant papers published in TSE, as one of the most popular journals, accounts for the largest proportion in 2018 and 2019; while another popular journal, IST, also makes up a large percentage in 2019 and 2020.

%nWe cabeen notice that the number of conference papers employing DL is over two times higher than that of journal papers.

%The 139 reviewed studies were published in various publication venues, including in 3 top journals and 3 top conferences, well-known and highly regarded in the field of software engineering. The prevalence of papers on predictive models in these conferences and journals indicates that the predictive model is considered a leading- edge approach to address various SE problems. Table 4 lists the number of papers published in each publication venue. ICSE (41) includes the highest number of primary study papers compared with other top conferences and journals, almost two times higher than ASE (22). In Table 4, TSE is the most popular journal containing almost 22% of the primary studies, followed by EMSE (21). 7 relevant studies were published in TOSEM. We can observe that the number of conference papers employing predictive models is higher than that of journal papers.
%We checked the distribution between different publication venues. Fig.3(a) shows the overall distribution of studies published in each venue, and Fig.3(b) gives the venue distribution per year. Fig.3(a) illustrates that 61% of publications appeared in conferences and symposiums while 39% were published in journals. As Fig.3(b) shows, from 2009 to 2016, most of the primary studies were published in conference and symposia proceedings. This trend has however changed, with the number of journal papers now increasing between 2014 and 2018, often outnumbering conference papers.

\subsection{Types of main contributions}

We summarized the main contribution of each primary study and then categorized these studies according to their main contributions into five categories, i.e., New technique or methodology, Tool, Empirical study, Case study, and User study. We gave the definition of each main contribution in Table~\ref{tab:definition of contribution}. The main contribution of 76\% primary studies was to build a novel DNN as their proposed new technique or methodology for dealing with various problems in different SE activities. 10\% of relevant studies concentrated on performing assessment and empirical studies for exploring the benefits of DL towards different SE aspects, such as research on the differences between ML and DL to solve certain SE tasks, the performance of using DL to mine software repositories, applying DL in testing, etc. The main contribution of 9\% was case studies. 2 primary studies (1\%) that both proposed a novel methodology and evaluated the novel methodology via a user study. Therefore, the main contribution of these two studies spans across two categories, i.e., New technique or methodology and user study.

% Fig. \ref{fig:type_contribution} shows the analysis result on the main contribution of primary studies.
\begin{table} 
\centering
\footnotesize
\caption{The definition of five main contributions in primary studies.}
\label{tab:definition of contribution}
 \vspace{-0.3cm}\begin{tabular}{p{4cm}p{8cm}}
  \toprule
 \textbf{Main contribution} & \textbf{Definition} \\
  \midrule
 New technique or methodology & The study provided a solid solution or developed a novel framework to address specific SE issues. \\
 Tool & The study implemented and published a new tool or tool demo targeting SE issues.\\
 Empirical study & The study collected primary data and performed a quantitative and qualitative analysis on the data to explore interesting findings. \\
 Case study & The study analyzed certain SE issues based on one or more specific cases. \\
 User study & The study conducted a survey to investigate the attitudes of different people (e.g., developers, practitioners, users, etc) towards SE issues. \\
 %practitioner
  \bottomrule
 \end{tabular}\vspace{-0.5cm}
\end{table}

% Only 2 primary studies (1\%) were to conduct a user study so as to investigate users' attitudes towards DL techniques in SE.

%\begin{figure}
%    \centering
%    \includegraphics[width=0.39\textwidth]{figure/types_of_contribution.png}
%    \caption{Types of main contributions}
%    \label{fig:type_contribution}
%\end{figure}

\begin{tcolorbox}[breakable,colback=gray!20,colframe=gray!35!black,title=Summary]
\small
 \begin{enumerate}
     \item DL has shown a booming trend in recent years
     \item Most of primary study papers were published between 2018 and 2020
     \item The number of conference papers employing DNNs for SE significantly exceeds that of journal papers.
     \item SANER is the conference venue publishing the most DL-based papers (22), while TSE and IST include the highest number of relevant papers among all journals (11).
     \item Most DL-based studies were only published in a few conference proceedings (e.g., SANER, ASE, ICSE, MSR) and journals (e.g., TSE, IST, JSS, and TOSEM).
     \item The main contribution of 75\% primary studies is to propose a novel methodology by applying various DL techniques, while only two primary studies performed a user study to better understand users' attitudes and experience toward various DNNs used for solving specific SE tasks.
 \end{enumerate}
\end{tcolorbox}

% show a booming trend, In the future
%2018 -- 2020
%conference > journal
%We can observe that the number of conference papers employing predictive models is higher than that of journal papers.
%SANER and TSE IST, 集中在几种期刊会议
%new mothod

\section{RQ2: What DL techniques are applied to support SE tasks?}

%In this section, we provide a detailed outline of diverse DL techniques, including classification and brief analysis of DNNs in terms of the model selection strategy.

\subsection{Classification of DNNs in SE}

\begin{table*} 
\newcommand{\tabincell}[2]{\begin{tabular}{@{}#1@{}}#2\end{tabular}}
\centering 	
\footnotesize
\caption{The number of various DNNs applied in per year.}
\label{tab:DNN models}
% \resizebox{\textwidth}{9.5cm}{
\vspace{-0.3cm}\begin{tabular} {p{0.15\columnwidth} p{0.1\columnwidth} p{0.2\columnwidth} l l l l l l l }
%\hline
\toprule
\textbf{Architecture} & \textbf{Family}  & \textbf{Model Name} & \textbf{2015} & \textbf{2016} & \textbf{2017} & \textbf{2018} & \textbf{2019} & \textbf{2020}  & \textbf{Total} \\ \midrule%\hline

\multirow{22}{*}{\tabincell{c}{Layered \\ architecture \\ (157)}} &\multirow{8}{*}{\tabincell{c}{RNN-based \\ model (72)}}
& RNN   & 1  &   & 3  & 7  & 10 & 2  &23 \\ %\cline{2-13}
&& RtNN &   & 1   &   & 1  &   &   & 2 \\ %\cline{2-13}
& & Bidirectional RNN (BRNN)&   &  &  &  & & 1   & 1  \\ %\cline{2-13}
&& LSTM  &   &  &  3   &  10  & 16    &   6  & 35   \\ %\cline{2-13}
&& Bi-LSTM   &   &   &   1  &  1   &   & 2   & 4  \\ %\cline{2-13}
&&Siamese LSTM  &   &   &   &   &   & 1  & 1   \\ %\cline{2-13}
&& GRU   &    &    &    & &  1  &  3    &4 \\ %\cline{2-13}
&& Bidirectional GRU  &  &  &   &   &    &  1    &1 \\
&& Recurrent Highway Network &  &   &  &  & 1  & &1   \\ \cline{2-10}
%\midrule%\hline

&\multirow{4}{*}{\tabincell{c}{CNN-based \\ model (48)}}
& CNN  & & 2   & 2   & 13   & 20   & 6   & 43 \\ %
&& Tree-based CNN (TBCNN) &   &  &  &   & 2  & 1   &3  \\ %\cline{2-13}
&& RCNN   &    &     &     &     &     & 1    & 1 \\ %\cline{2-13}
& & Deep Residual Network  &  &  &   &   &  & 1   &1 \\ \cline{2-10}
 %\midrule%\hline

& \multirow{4}{*}{\tabincell{c}{FNN-based \\ model (25)}}
 & FNN  &    & 3  &1 &  8 & 7 & 3  &22 \\ %
& & RBFNN & 1   & & & &   &   &1 \\ %
&& Deep Sparse FNN  &    &   &    &   1   &  & & 1  \\
&& Deep MLP & &    &    & &  1&    &1\\
\cline{2-10}
 %\midrule%\hline

& \multirow{2}{*}{\tabincell{c}{GNN-based \\ model (6)}}
 & GGNN  &   &   & &   & 4 & 1  &5 \\ %
& & Graph Matching Network (GMN) &   & & & &   & 1  &1 \\
\cline{2-10}
 %\midrule%\hline

& \multirow{3}{*}{\tabincell{c}{Tailored \\ model (4)}}
%  & combined DNN  &    & 1  & &   &  & 1  &2 \\ %
& Deep Beliefe Network (DBN) &    &1 & &1 &   &   &2 \\
&& HAN  &    &   &    &     & 1 & & 1  \\
&& Deep Forest & &    &   & &  1&  &1\\
\midrule%\hline

\multirow{6}{*}{\tabincell{c}{Encoder-Decoder \\ (15)}} &\multirow{2}{*}{\tabincell{c}{RNN-based \\ model (12)}}
& RNN   &   &  1 &   & 1  & 6 &   &8 \\
&& LSTM  &   &  &   &  2  &  & 2  & 4 \\
%\cline{2-10}

&\multirow{2}{*}{\tabincell{c}{CNN-based \\ model (1)}} & CNN  & &    &   &   & 1  &   & 1 \\ %
&&  & &    &   & &  &  &\\
%\cline{2-10}

&\multirow{2}{*}{\tabincell{c}{FNN-based \\ model (2)}}
& FNN  & &    &   & 1  &   & 1  & 2 \\ %
&&  & &    &   & &  &  &\\
\midrule

\multirow{6}{*}{\tabincell{c}{AutoEncoder \\ (7)}} &\multirow{2}{*}{\tabincell{c}{RNN-based \\model (1)}}
& GRU   &   &   &   &   &  & 1  &1 \\
&&  & &    &   & &  &  &\\
%\cline{2-10}

&\multirow{2}{*}{\tabincell{c}{CNN-based \\ model (1)}}
& CNN  & &    &   &   &   &  1 & 1 \\ %
&&  & &    &   & &  &  &\\
%\cline{2-10}

&\multirow{2}{*}{\tabincell{c}{FNN-based \\  model (5)}}
& FNN  & &   & 2  & 1  &  2 &   & 5 \\ %
&&  & &    &   & &  &  &\\

\bottomrule
\end{tabular}\vspace{-0.5cm}
\end{table*}

Many sorts of DNNs have been proposed, and certain neural network architectures contain diverse DNNs with different implementations.  For instance, although LSTM and GRU are considered two different DNNs, they are both RNNs. We categorized DL-based models according to their architecture and different DNNs used. We classified the architecture of various DNNs into 3 categories: the layered architecture, AutoEncoder (AE), and Encoder-Decoder \cite{deng2014tutorial, nayak2020effective}. We provided a detailed classification of DNNs into five categories, i.e., RNN, CNN, FNN, GNN, and tailored DNN models, where tailored DNNs include the DNNs not often used in SE, e.g., DBN, HAN, etc. Table ~\ref{tab:DNN models} shows the variety of different DNNs, and also lists the number of times these models have been applied in SE.

As can be seen from Table ~\ref{tab:DNN models} that compared Encoder-Decoder and AutoEncoder (AE) architectures, layered based DNNs are the most popular and widely used architecture. In the layered architecture, 72 primary studies used nine different kinds of RNN-based models to solve practical SE issues, where LSTM is the most often applied RNN-based model (35), followed by standard RNN (23). The variants of LSTM, such as GRU and Bi-LSTM, are often adopted by researchers in multiple research directions. 48 primary studies employed CNN-based models, where almost 90\% of studies employed CNN. FNN-based model is the third most frequently used family with 25 studies using FNNs, followed by GNN-based models and tailored models. There are 24 combined DNNs were proposed in tailored models.

15 primary studies leveraged different types of DNNs following the Encoder-Decoder architecture, where RNNs were used in 12 studies, which is much higher than the number of other models used, i.e., CNN and FNN. In the last architecture, over 70\% of studies used FNN-based AEs as their proposed novel approaches; only 2 studies selected GRU and CNN to construct AEs respectively.
%Furthermore, we can observe that a great number of DL-based models were proposed between 2018 and 2020, once again indicating the prevalence of DL techniques in recent years.

\subsection{DL technique selection strategy}

Since heterogeneous DNNs have been used for SE tasks, selecting and employing the most suitable network is a crucial factor. We scanned the relevant sections of DL technique selection in all of the selected primary studies and classified the extracted rationale into three categories.

%\begin{enumerate}
\textbf{Characteristic-based selection strategy ($S1$)}: The studies justified the selected techniques based on their characteristics to overcome the obstacles associated with a specific SE issue \cite{chen2018ui, gu2016deep, zhou2020improving, enicser2020virtualization}. For instance, most of the seq2seq models were built by using RNN-based models thanks to their strong ability to analyze the sequence data.

 \textbf{Selection based on prior studies ($S2$)}: Some researchers  determined the most suitable DNN used in their studies by referring to the relevant DL techniques in the related work \cite{hussain2020codegru, zhang2019novel, cambronero2019deep}. For instance, due to the good performance of CNN in the field of image processing, most studies selected CNN as the first option when the dataset contains images.

 \textbf{Using multiple feasible DNNs ($S3$)}: Though not providing any explicit rationale, some studies designed experiments for technique comparisons that demonstrated that the selected algorithms performed better than other methods. For example, some studies often selected a set of DNNs in the same SE tasks to compare their performance and picked up the best one \cite{alahmadi2020code, tian2020bvdetector, dq2019bilateral}.
%\end{enumerate}

% \begin{figure}
%     \centering
%     \includegraphics[width=0.4\textwidth]{figure/selection.png}
%     \caption{DL-based model selection strategies}
%     \label{fig:selection}
% \end{figure}

%FIg.~\ref{fig:selection} describes the selection trend of DNNs in all primary studies. T
We noticed that the most commonly selection strategy is \textbf{$S1$} (i.e., Characteristic-based selection strategy), accounting for \textbf{69\%}, nearly 3 times that of\textbf{ $S2$ (25\%)}. Only \textbf{6\%} of primary studies adopt \textbf{$S3$} to select their suitable DL algorithms. %\hl{One potential reason may be that adopting the last selection strategy not only significantly increases researchers' workload but may unable to find the best one.}

\begin{tcolorbox}[breakable,colback=gray!20,colframe=gray!35!black,title=Summary]
\small
 \begin{enumerate}
     \item There are 30 different DNNs used in the selected primary studies.
     \item We used a classification of DL-based algorithms from two perspectives, i.e., their architectures and the families to which they belong. The architecture can be classified into three types: Layered architecture, Encoder-Decoder, and AutoEncoder (AE); the family can be classified into five categories: RNN-based, CNN-based, FNN-based, GNN-based, and Tailored models.
     \item Compared with Encoder-Decoder and AE, the layered architecture of DNNs is by far the most popular option in SE.
     \item Four specific DNNs are used in more than 20 primary studies, i.e., CNN (43), LSTM (35), RNN (23), and FNN (22), and each of them has several variants that are also often used in many SE tasks.
     \item We summarized three types of DNN-based model selection strategies. The majority of studies adopted $S1$ to select suitable DL algorithms. Only 6\% of primary studies used $S3$ as the model selection strategy due to the heavy workload brought by $S3$.
 \end{enumerate}
\end{tcolorbox}

%The core idea is to
%the key innovation of
%Experimental results based on 2 benchmark detasets and 6 long-lived open-source software projects show that
%the perofrmance of XXX is superior to state-of-the-art baseline methods.
%the final/experimental/retrieval results
%The results show/suggest/confirm/indicate /  significantly better effort-aware ranking effectiveness than
%proposed, present, introduced, pinpoint,
%evaluate, measure
% satisfiable, shallow features
%previous work
%use, leverage, exploit, apply, employ, adopt, take advantage of
%is equipped with the ability
%heterogeneous, various, diverse,

\section{RQ3: What key factors contribute to difficulty when training DNNs in SE?}

%After researchers adopted different model selection strategies to determine the most suitable DNNs in model construction phase, they also pay much attention on training phase since many key techniques are involved in this phase, which having great impact on the performance of their proposed models.

Since analyzing a DL architecture can provide a lot of insight, we investigated the construction process of a DL framework from three aspects:  techniques used in data processing, model optimization, evaluation, and the accessibility of primary studies.

\subsection{How were datasets collected, processed, and used?}

Data is one of the most important roles in the training phase. Unsuitable datasets can result in failed approaches or tools with the low performance. We focused on the data used in primary studies and conducted a comprehensive analysis on the steps of data collection, data processing, and data application.

\subsubsection{What were the sources of datasets used for training DNNs?}
\begin{figure}%[htbp]
\centering
\subfigure[The sources of DL-related dataets.]{
\centering
\begin{minipage}[t]{0.38\textwidth}
\includegraphics[width=1\textwidth]{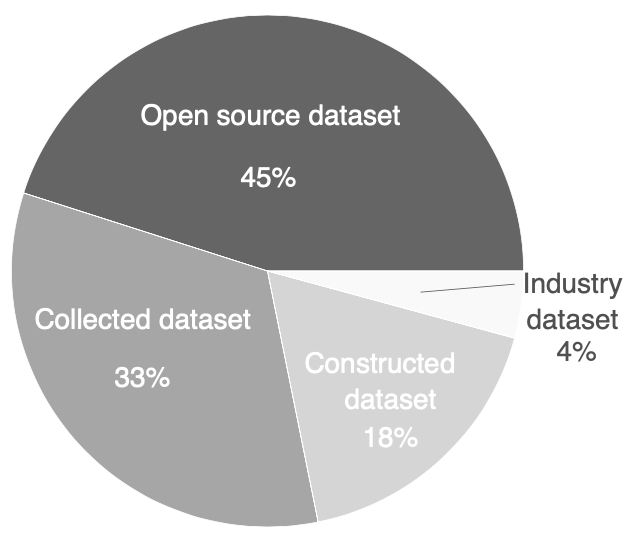}
\end{minipage}
}
%\hfill
\subfigure[The sources of collected datasets.]{
%\centering
\begin{minipage}[t]{0.38\textwidth}
\includegraphics[width=0.86\textwidth]{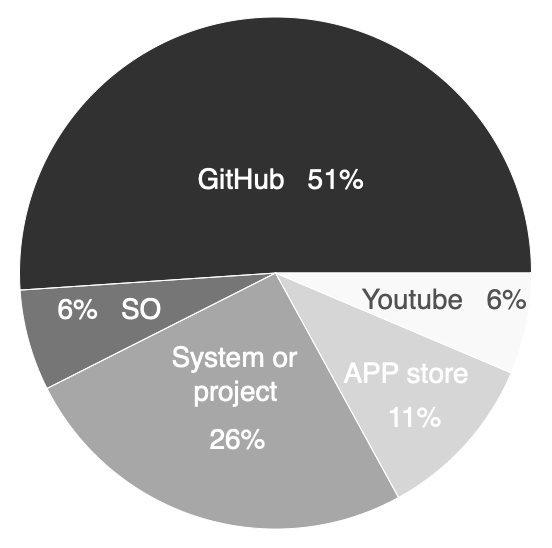}
\end{minipage}
}
\vspace{-0.3cm}\caption{The source of datasets used in primary study papers}\vspace{-0.6cm}
\label{fig:dataset_source}
\end{figure}

 Fig.~\ref{fig:dataset_source} shows the sources of datasets in the selected primary studies. It can be seen that 45\% of primary studies trained DNNs by using an open-source dataset. One reason for choosing an open-source dataset is that studies are willing to pick up these datasets to evaluate the effectiveness of proposed DL-based approaches due to the existence of widely accepted datasets in certain SE issues (e.g., code clone detection, software effort/cost prediction, etc). Others are used because the datasets were applied in related previous work. Due to the lack of available and suitable datasets, 18\% of primary studies constructed new datasets. Real-world datasets from industry are only used by 4\% of studies.

33\% of studies performed a series of experiments on large-scale datasets so as to verify the scalability and robustness of their models. To achieve this, many studies collected multiple small datasets from different sources. Fig.~\ref{fig:dataset_source}(b) describes the sources of collected datasets. As tens of thousands of developers contribute to GitHub community by uploading source code of their software artifacts, GitHub has become the most frequently used source of collected data (51\%). 26\% of studies collected their datasets from different systems and projects. For instance, Deshmukh et al. \cite{deshmukh2017towards} collected bug reports from several bug tracking systems, i.e., issue tracking systems of Open Office, Eclipse, and Net Beans projects, as datasets to build a DL-based model for duplicated bug detection. The app store is the third-largest source (11\%), followed by Stack Overflow (SO) and Youtube.

\subsubsection{What are the types of SE datasets used in prior DL studies?} The datasets used in primary studies are of various data types. It is essential to analyze data types of datasets since the relationship between the type of implicit feature being extracted and the architecture has a dominating influence on model selection. We summarize the data types in primary studies and interpreted how data types determine the choice of DNNs.

\begin{table*} \vspace{-0.2cm}
\newcommand{\tabincell}[2]{\begin{tabular}{@{}#1@{}}#2\end{tabular}}
\centering 	
\footnotesize
\caption{Data types of datasets involved in primary studies.}
\label{tab:data types in datasets}
% \resizebox{\textwidth}{9.5cm}{
\vspace{-0.3cm}\begin{tabular} {ccccccccc }
%\hline
\toprule
\textbf{Family}  & \textbf{Data types} & \textbf{\#Studies}  & \textbf{Total} \\ \midrule%\hline

\multirow{11}{*}{\tabincell{c} Code-based data types}
& Source code   &61 & \\ %\cline{2-13}
& Software/code metric  & 8 & \\ %\cline{2-13}
& Code comment & 7 &  \\ %\cline{2-13}
& Defects & 7 &   \\ %\cline{2-13}
& Test case & 6 &  \\ %\cline{2-13}
& program screencasts  & 4 & \\ %\cline{2-13} %graph
& UI images  & 4 & \\ %\cline{2-13} %graph
& Code change & 2 &   \\ %\cline{2-13}
& Code annotation & 2 & \\ %\cline{2-13}
& Pull-requests & 2 & \\ %graph
& Patch & 1 & \multirow{-11}{*}{\tabincell{c} 104} \\
\midrule%\hline

\multirow{9}{*}{\tabincell{c} Text-based data types}
& Bug report  & 9 & \\ %\cline{2-13}
% & Questions and answers in SO & 6 & \\
% & Tags in SO & 5 & \\
& Requirement documentation  & 4 & \\ %\cline{2-13}
% & Issues and commits in GitHub & 4 & \\
% & Pull requests in GitHub & 2 & \\
& configuration documentation & 2 &  \\ %\cline{2-13}
& APP description & 2 &   \\ %\cline{2-13}
& Software version information & 2 &  \\ %\cline{2-13}
& Design documentation & 1 &   \\ %\cline{2-13}
& Log information & 1 & \\ %\cline{2-13}
& Certification & 1 & \\ %\cline{2-13}
& Protocol message & 1 & \\ %\cline{2-13}
& Patch & 1 & \multirow{-9}{*}{\tabincell{c} 23} \\
\midrule%\hline

\multirow{4}{*}{\tabincell{c} Software repository-based data types}
& Q\&A in SO  & 6 & \\ %\cline{2-13}
& Tags in SO  & 5 & \\ %\cline{2-13}
& Issues and commits & 4 &  \\ %\cline{2-13}
& Pull-requests & 2 &  \multirow{-4}{*}{\tabincell{c} 17} \\ %\cline{2-13}
\midrule%\hline

% \multirow{3}{*}{\tabincell{c} Graph-based data types}
% & program screencasts  & 4 & \\ %\cline{2-13}
% & UI images  & 4 & \\ %\cline{2-13}
% & Pull-requests & 2 &  \multirow{-3}{*}{\tabincell{c} 9} \\
% \midrule%\hline

\multirow{3}{*}{\tabincell{c} User-based data types}
& User behavior  & 3 & \\ %\cline{2-13}
& User review  & 1 & \\ %\cline{2-13}
& Interaction traces & 1 &  \multirow{-3}{*}{\tabincell{c} 5} \\
\bottomrule
\end{tabular}\vspace{-0.6cm}
\end{table*}

We classified the data types of used datasets into four categories -- code-based, text-based, software repository-based, and user-based data types. Table~\ref{tab:data types in datasets} describes specific data in each data type. 104 primary studies collected data from source code, and most of these studies used source code directly in some important SE activities, such as software testing and maintenance. Datasets containing various metrics were employed in 8 relevant studies, followed by code comments, defects (7), and test cases (6). Whereas few studies focused on analyzing the code annotation, pull-request, and patches. 8 primary studies used a multitude of screencasts as their datasets, where 4 studies selected program screencasts to analyze developers' behavior and 4 studies researched UI images for improving the quality of APPs.

Text-based data types were the second most popular, including 13 different kinds of documentation. Bug report and requirements documentation are the two most commonly applied text-based data types in primary studies. Some types rarely appeared, such as logs, certifications, design documentation, etc.

%Since a number of studies concentrated on mining software repositories for achieving useful patterns or information, massive information crawled from SO (e.g., questions and tags) and GitHub (e.g., issues, commits and pull-requests) as different kinds of datasets was used in around 12\% of studies.

Since software repositories, especially SO and GitHub, contain a lot of useful patterns or information, we classified the type of the information collected from these repositories into $'Software$ $repository-based$ $data$ $types'$. 12\% of studies concentrated on obtaining and learning useful information and patterns by crawling related contents from SO (e.g., Q\&A (questions and answers) and tags) and GitHub (e.g., issues, commits and, pull-requests).

User-based data generally contains a great deal of user information, which can promote developers to better comprehend user needs and behavior targeting different applications. Only 5 studies adopted user-based data types (i.e., user behavior, review, and interactions) to solve relevant SE tasks.

\subsubsection{What input forms were datasets transformed into when training DNNs?}

The inputs of DNNs need to be various forms of vectors. We found two techniques were often used to transform different source data types into vectors: One-hot encoding and Word2vec. Only 5 studies produced the input of their models by adopting the One-hot technique. We described input forms using 5 categories referring to their data types.

\textbf{Token-based input}: Since some studies treated source code as text, they used a simple program analysis technique to generate code tokens into sequences and transformed tokens into vectors as the input of their DL-based models. A token-based input form can be applied to source code and text-based data when processing related datasets.

\textbf{Tree/graph-based input}: To better comprehend the structure of source code, several studies convert source code into Abstract Syntax Trees (AST) or Control Flow Graphs (CFGs), and then generate vector sequences by traversing the nodes in each tree or graph.

\textbf{Feature/metric-based input}: For analyzing the characteristics of software artifacts, some studies applied datasets consisting of features or metrics extracted from different products, and thus the input form of the models proposed in these studies is software feature/metric-based vectors.

\textbf{Pixel-based input}: Some studies used datasets containing a large number of images and screencasts, e.g., program screencasts, UI images, etc. When preprocessing these datasets, they often broke down screencasts into pixels as an effective input form, for analyzing graph-based datasets in different SE tasks, such as bug detection, code extraction, etc.

\textbf{Combined input}: Many studies combined two or more data types extracted from software products to build comprehensive datasets with more information for enhancing the quality and accuracy of proposed models. For instance, Leclair et al. \cite{leclair2019neural} proposed a novel approach for generating summaries of programs not only by analyzing their source code but also their code comments.

\begin{table*} 
\newcommand{\tabincell}[2]{\begin{tabular}{@{}#1@{}}#2\end{tabular}}
\centering 	
\footnotesize
\caption{The various input forms of DL-based models proposed in primary studies.}
\label{tab:input forms in datasets}
% \resizebox{\textwidth}{9.5cm}{
\vspace{-0.3cm}\begin{tabular} {ccccccccc }
%\hline
\toprule
\textbf{Family}  & \textbf{Input forms} & \textbf{\#Studies}  & \textbf{Total} \\
\midrule%\hline

\multirow{2}{*}{\tabincell{c} Token-based input}
& Code in tokens  & 17 & \\ %\cline{2-13}
& Text in tokens  & 34 & \\
& Code and text in tokens & 13 & \multirow{-2}{*}{\tabincell{c} 64} \\

\midrule%\hline
\multirow{2}{*}{\tabincell{c} Tree/graph-based input}
& Code in tree structure  & 25 & \\ %\cline{2-13}
& Code in graph structure  & 4 & \multirow{-2}{*}{\tabincell{c} 29} \\
\midrule%\hline

Feature-based input & feature/metric & 33 & 33 \\ %\cline{2-13}
\midrule%\hline

Pixel-based input & pixel & 9 & 9 \\
\midrule%\hline

\multirow{3}{*}{\tabincell{c} Hybrid input}
& Code in tree structure + text in token  & 4 & \\ %\cline{2-13}
& Code features + text in token  & 2 & \\ %\cline{2-13}
& Code in tree structure + features & 1 &  \multirow{-3}{*}{\tabincell{c} 7} \\
\bottomrule
\end{tabular}\vspace{-0.5cm}
\end{table*}

Table~\ref{tab:input forms in datasets} depicts the input formats of DL-based models. We can see that over 45\% of studies transformed data (i.e., source code and various documentations) into the token-based input form (64), where 17 studies considered source code as texts and thus converted code into token sequences as the input of models. 13 studies used both source code and text-based materials and also constructed a token-based data structure as the input form of their proposed models. 25 studies utilize tree-based input form to analyze the source code, and only 4 studies transform source code into a graph-based structure for extracting essential information. 33 studies adopted embedding techniques to generate feature-based vectors.
Furthermore, 8 studies using image-based datasets split screencasts into pixels as the basic unit of the input form. Only 7 studies processed datasets into multiple forms.

%\subsubsection{which data preprocessing techniques are often used for training Dl-based models?}

\subsection{What techniques were used to optimize and evaluate DL-based models in SE?}

In the training phase, developers attempt to optimize the models in different ways for achieving good performance. In this section, we summarized the information describing the optimization methods and evaluation process, and performed an analysis on key techniques.

\subsubsection{What learning algorithms are used in order to optimize the models?}

The performance of DL-based models is dependent on selected optimization methods, which can systematically adjust the parameters of the DNN as training progresses.

\begin{figure}
    \centering
    \includegraphics[width=0.8\textwidth]{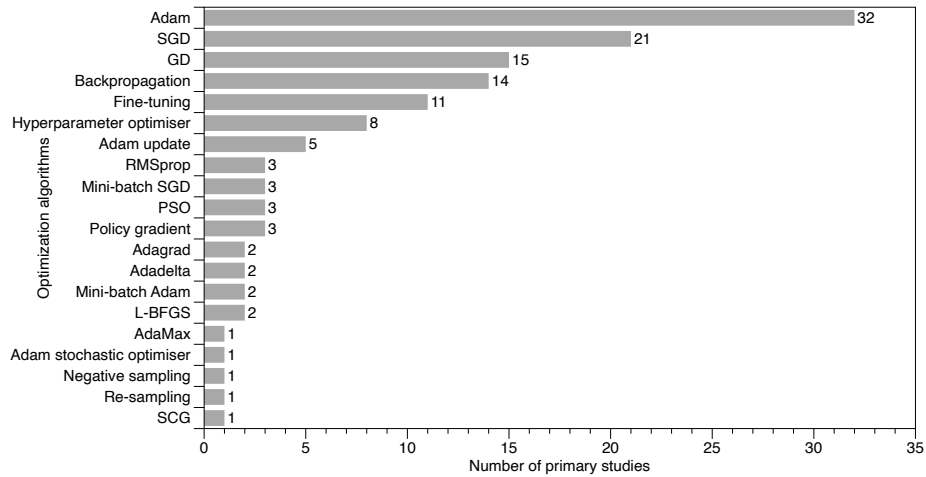}
   \vspace{-0.3cm} \caption{Various optimization algorithms used in primary studies.}
    \label{fig:optimization methods}\vspace{-0.5cm}
\end{figure}

Out of the 142 studies analyzed, 131 identified the specific optimization method, but 11 studies did not mention what optimizers were used to adjust parameters in their work. Fig.~\ref{fig:optimization methods} illustrated the frequency of the use of 20 optimization methods used in all primary studies. We see that 6 optimizers were used in no less than 5 studies, where Adam optimizer is the most commonly used optimization method. Stochastic gradient descent (SGD) and gradient descent (GD) are also popular optimizers, which were used by 21 and 15 studies respectively, followed by back-propagation (14) and fine-tuning (11). Besides, some optimization methods are not often used, such as Adagrad and Adadelta.

%Besides, although some optimization methods are well-known, there are not many studies using these methods, such as Adagrad and Adadelta.

\subsubsection{What methods were used to alleviate the impact of Overfitting?}

One major problem associated with applying any type of learning algorithm is overfitting. Overfitting is the phenomenon of a DNN learning to fit the noise in the training data extremely well, yet not being able to generalize to unseen data, which is not a good approximation of the true target function that the algorithm is looking forward to learn. We describe 9 general ways to combat overfitting \cite{hawkins2004problem, srivastava2014dropout} by considering 3 aspects: data processing, model construction as well as model training, and then analyzed the usage distribution of these methods in relevant studies.

%\begin{enumerate}
   \textbf{Cross-validation}: A cross-validation algorithm can split a dataset into k groups (k-fold cross-validation). Researchers often leave one group to be the validation set and the others as the training set. This process will be repeated until each group has been used as the validation set. Since a remaining subset of data is new towards the training process of DL-based models, the algorithm can only rely on the learned knowledge from other groups of data to predict the results of the remaining subset, preventing overfitting.

    \textbf{Feature selection}:  Overfitting can be prevented by selecting several of the most essential features for training DL-based models can effectively avoid overfitting. Therefore, researchers can pick up some key features by using feature selection methods, train individual models for these features, and evaluate the generalization capabilities of models. For instance, Pooling is a typical technique to prevent overfitting since pooling can reserve main features while reducing the number of parameters and the amount of computation, and improve the generalization ability of the model.

    \textbf{Regularization}: Regularization is a technique to constrain the network from learning a model that is too complex, which therefore can avoid overfitting. A penalty term would be added in the cost function to push the estimated coefficients towards zero by applying L1 or L2 regularization.

   \textbf{Dropout}: By applying dropout, a form of regularization, to the layers of DNNs, a part of neurons were ignored with a set probability. Therefore, dropout can reduce interdependent learning among units to avoid overfitting.

   \textbf{Data augmentation}: A larger dataset can reduce the chance of overfitting. Data augmentation is a good way to artificially increase the size of our dataset for improving the performance of a DNN  when the scale of data was constrained due to difficult to gather more data. For example, many studies performed various image transformations to the image dataset (e.g., flipping, rotating, rescaling, shifting) for enlarging data size in the image classification task.

    \textbf{Early stopping}: Early stopping is an effective method for avoiding overfitting by truncating the number of iterations, that is, stopping iterations before DL-based models converge on the training dataset to eliminate the impact on overfitting.

     \textbf{Data balancing}: With imbalanced data, DL-based models are likely to occur the overfitting problem since models will learn imbalanced knowledge with a disproportionate ratio of observations in each class. Using some data balancing techniques can effectively alleviate the impact on models' performance caused by overfitting.

     \textbf{Ensembling}: Ensembles are a set of machine learning methods for combining predictions from multiple separate models. For instance, Bagging as an ensemble learner can reduce the chance of overfitting complex models by training a large number of "strong" learners in parallel without restriction.
     %\textbf{Others}: There are also some other methods to solve the overfitting problem, such as some proposed new algorithms.
%\end{enumerate}

\begin{figure}
    \centering
    \includegraphics[width=0.7\textwidth]{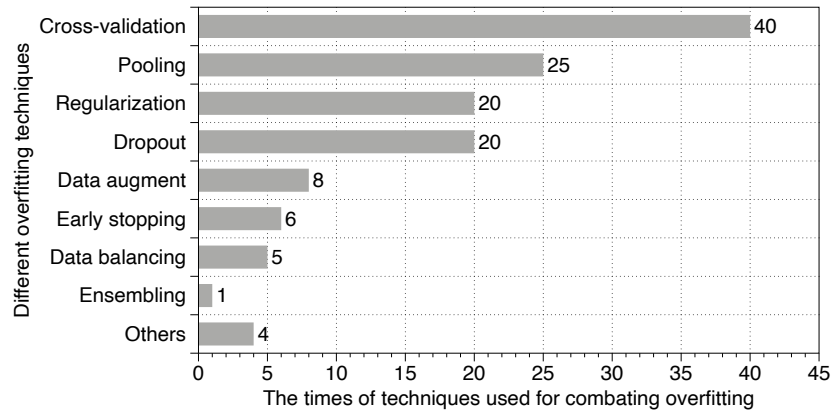}
    \vspace{-0.3cm}\caption{The distribution of various overfitting techniques used in primary studies}
    \label{fig:overfitting}\vspace{-0.5cm}
\end{figure}

Fig.~\ref{fig:overfitting} illustrates the distribution of the techniques used for combating overfitting problems. Cross-validation has been used frequently among the selected studies to prevent overfitting; it is used in 40 studies, followed by pooling (25). Regularization and dropout are the third most popular techniques used in 20 studies. There are 8 studies that prevent overfitting by enlarging the scale of data, such as using a large-scale dataset, combining multiple datasets, and using different data augmentation techniques. 6 studies used early stopping and 5 selected data balancing to combat the overfitting problem. Ensembling is the least frequently used one (1) compared with other techniques (1). Furthermore, among all primary studies, 4 studies used several new algorithms proposed by some studies to solve overfitting.
We analyzed which factors may have an impact on the overfitting technique selection. We noticed that the techniques used for combating overfitting have no strong association with either data types or input forms. However, there is a special relationship between model selection and these techniques. Most of the studies that adopted CNNs to address specific SE tasks selected pooling as their first choice for preventing the overfitting problem.

\subsubsection{What measures are used to evaluate DL-based models?}

Accessing appropriate benchmarks is a crucial part of evaluating any DL-based models in SE. We also explored the frequent metrics used to measure the performance of DL-based models applied to respective SE tasks.

%Understanding baseline approaches and accessing appropriate benchmarks is a crucial part of evaluating any DL based approach in SE. However, we also explored the common metrics used to measure the performance of DL models applied to their respective SE tasks. In many instances, the metrics chosen to analyze the model are common to the type of learning. Therefore, many of the supervised learning methods have metrics that analyze the resulting hypothesis, such as the accuracy, true positives, F1 measure, precision and recall. These metrics are used to compare the supervised learning algorithms with the outputs representing the target hypothesis. Intuitively, the type of metric chosen to evaluate the DL based approach is dependent upon the data type and architecture employed by the approach.

\begin{table*}[!htbp]
\newcommand{\tabincell}[2]{\begin{tabular}{@{}#1@{}}#2\end{tabular}}
\centering 	
\footnotesize
\caption{Metrics used for evaluation.}
\label{tab:metrics}
% \resizebox{\textwidth}{9.5cm}{
\vspace{-0.3cm}\begin{tabular} {p{6cm}ccccccc }
%\hline
\toprule
\textbf{Metrics} & \textbf{\#Studies} \\
\midrule%\hline
Precision@k & 69 \\
Recall@k & 59 \\
F1@k & 53 \\
Accuracy & 26 \\
Mean Reciprocal Rank (MRR) & 15 \\
BLEU & 13 \\
Running time & 13 \\
AUC & 11 \\
Mean Average Precision (MAP) & 7 \\
Matthews Correlation Coefficient (MCC) & 5 \\
P-value & 4 \\
METEOR & 4 \\
ROC & 4 \\
ROUGE & 4 \\
Mean Absolute Error (MAE) & 4 \\
SuccessRate@k & 3 \\
Cliff’s Delta & 3 \\
Coverage & 3 \\
Bal (Balance) & 3 \\
Standardized Accuracy (SA) & 3 \\
Others & 11 \\
 \bottomrule
\end{tabular}\vspace{-0.5cm}
\end{table*}

Table~\ref{tab:metrics} summarizes the commonly used evaluation metrics in the primary studies, used in no less than 3 studies. Precision, recall, F1-measure, and accuracy are widely accepted metrics for evaluating the performance of DL-based models. Some studies adopted MRR and BLEU as evaluation metrics in their work, potentially indicating that many studies focused on addressing ranking and translation tasks by training various DNNs. Another interesting observation from Table~\ref{tab:metrics} is that running time is selected as a performance indicator by a set of studies, which does not occur frequently when using non-learning techniques. This is because that learning algorithms, especially DNNs, require more time during their construction, training, and testing phases due to the high complexity of these networks (e.g., numerous types of layers, a great many neurons, and different optimization methods). Also, almost half of metrics are not commonly used in relevant studies, which are only used in 3 or 4 studies (e.g., P-value, ROC, ROUGE, Coverage, Balance, etc.) and thus these metrics can reflect their respective characteristics of different SE tasks.

\subsection{Accessibility of DL-based models used in primary studies.}

%The replicability and reproducibility of DL applications have a great impact on the transfer of research results into industry practices. According to the ACM policy on artifact review and badging [60], replicability refers to the ability of an independent group to obtain the same result using the author’s own artifacts. Likewise, reproducibility is the act of obtaining the same result without the use of original artifacts (but generally using the same methods). Reproducibility is clearly the ultimate goal, but replicability is an intermediate step to promote practices. Somewhat unfortunately, according to Fu and Menzies [19], it is hard to replicate or reproduce DL applications from SE research due to the nondisclosure of datasets and source code.

We checked whether the source code of DL-based models is accessible for supporting replicability and reproducibility. \textbf{53} studies provided the replication packages of their DL-based models, only accounting for \textbf{37.3\%} of all primary studies. \textbf{89} studies proposed novel DL-based models without publicly available source code, making it difficult for other researchers to reproduce their results; some of these studies only disclosed their datasets.
Based on this observation, obtaining open-source code of DNNs is still one of the challenges in SE because many factors may result in never realizing the replicability and reproducibility of DL application, e.g., data accessibility, source code accessibility, different data preprocessing techniques, optimization methods, and model selection methods. Therefore, we recommend future DL studies to release replication packages.

\begin{tcolorbox}[breakable,colback=gray!20,colframe=gray!35!black,title=Summary]
\small
 \begin{enumerate}
     \item Most datasets are available online and 33\% of datasets consist of multiple small-scale ones collected from GitHub, software systems and projects, and some software repositories.
     \item 5 different data types are used in the primary studies, i.e., code-based, text-based, software repository-based, graph-based, and user-based data types, where code-based and text-based types are the two main data types being used in 82.3\% of primary studies.
     \item Most studies parse source code into token, tree, or graph structures, or extract features from programs. When the raw datasets are documentation, studies would convert them into token-based vectors as the input form of their models.
     \item We observed that Adam is the most popular optimization algorithm being used in 32 studies, followed by SGD and GD, and several variants of Adam are still commonly used in SE. There are also some well-known optimization algorithms used in primary studies, such as back propagation, fin-tuning, and hyperparameter optimizer.
     \item 9 different ways are used for combating the overfitting problem. 4 techniques were widely used -- cross-validation, feature selection, regularization, and dropout. We found that studies applying CNNs would choose the pooling method to prevent overfitting with high probability, and  data type and input form does not influence technique selection.
     \item There are over 20 different metrics used to verify the performance of DL-based models. Precision, recall, F1-measure, and accuracy were used commonly in primary studies.
     \item Only 53 studies provided a public link of their models in their papers and yet 62.7\% of proposed models are difficult to be reproduced since the source code is unavailable.
 \end{enumerate}
\end{tcolorbox}

\section{RQ4: What types of SE tasks and which SE phases have been facilitated by DL-based approaches?}

%As researchers and practitioners have paid more attention to DL due to its unique structure, many DL models have been introduced, improving the performance of various SE tasks in different SE activities, such as defect prediction, source code summarization, code search, and code review.
In this section, we first categorise a variety of SE tasks into six SE activities referring to Software Engineering Body of Knowledge \cite{bourque2014guide}, i.e., software requirements, software design, software implementation, software testing and debugging, software maintenance, and software management. We then analyze the distribution of DL-based studies for different SE activities. We present a short description of each primary study, including the specific SE issue each study focused on, which and how DL techniques are used, and the performance of each DL model used.

\subsection{Distribution of DL techniques in different SE activities}

\begin{figure} 
\centering
\label{fig:SE}
\begin{minipage}[t]{0.38\textwidth}
\centering
\includegraphics[width=1.12\textwidth]{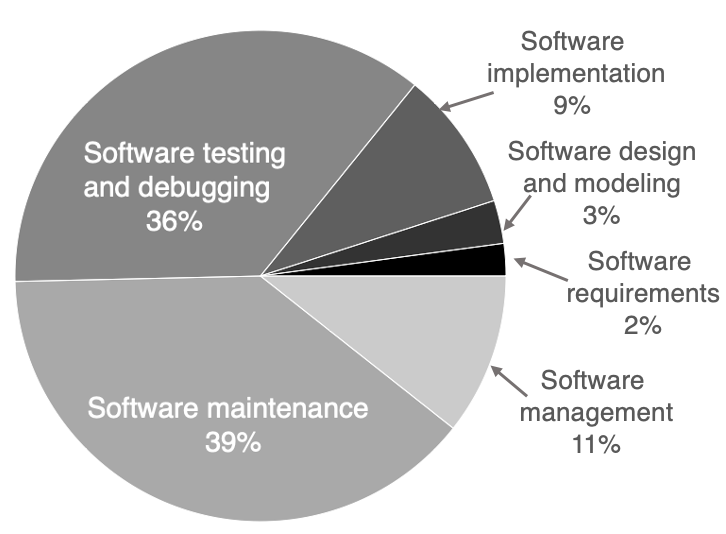}
 \caption{The distribution of DL techniques in Different SE activities.}
\label{fig:SE_activities}
\end{minipage}
\begin{minipage}[t]{0.38\textwidth}
\centering
\includegraphics[width=0.8\textwidth]{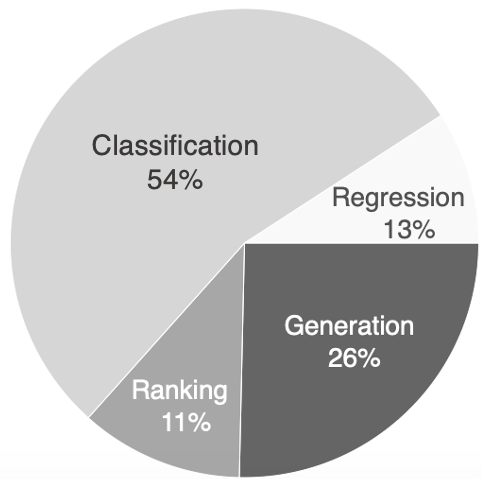}
 \caption{The classification of primary studies.}
\label{fig:studies_classification}
\end{minipage}\vspace{-0.3cm}
\end{figure}

We analysed which SE activities and specific SE tasks each selected primary study tried to solve. As shown in Fig.~\ref{fig:SE_activities}, the largest number of primary studies focused on addressing SE issues in software maintenance (39\%). 36\% of studies researched software testing and debugging. Software management was the topic of 11\% of primary studies, followed by software implementation (9\%). Software design and modeling (3\%) and software requirements (2\%) are addressed in very few studies.

We classified all primary studies into four categories based on the types of their SE tasks, i.e., the regression task, classification task, ranking task, and generation task. Fig.~\ref{fig:studies_classification} describes the distribution of different task types where DL techniques were applied. Classification and generation tasks account for almost 80\% of primary studies, where classification is the most frequent task (54\%).  13\% of studies belong to the regression task and the output of their proposed models is a prediction value, such as effort cost prediction. In SE, some studies adopted DL to concentrate on a ranking task, accounting for 11\% of all studies.

\begin{table*} 
\newcommand{\tabincell}[2]{\begin{tabular}{@{}#1@{}}#2\end{tabular}}
\centering 	
\footnotesize
\caption{The specific research topics where DL techniques are often applied.}
\label{tab:table_topics}
% \resizebox{\textwidth}{9.5cm}{
\vspace{-0.3cm}\begin{tabular} {ccccccccc}
%\hline
\toprule
\textbf{SE activities} & \textbf{specific research topics} & \textbf{\#Studies}  & \textbf{Total} \\
\midrule%\hline

\multirow{1}{*}{\tabincell{c} Software design}
% & Code in tokens  & 17 & \\ %\cline{2-13}
% & Text in tokens  & 34 & \\
& Source code representation & 5 & \multirow{-1}{*}{\tabincell{c} 5} \\
\midrule%\hline

\multirow{2}{*}{\tabincell{c} Software implementation}
%& Defect prediction  & 11 & \\ %\cline{2-13}
& Code search  & 5 & \\
& Code programming & 4 & \multirow{-2}{*}{\tabincell{c} 9} \\
\midrule%\hline

\multirow{6}{*}{\tabincell{c} Software testing and debugging}
& Defect prediction  & 11 & \\ %\cline{2-13}
& Bug localization  & 7 & \\
& Application testing & 7 & \\
& Program analysis & 5 & \\
& Test case generation & 4 &\\
& Reverse execution & 3 & \multirow{-6}{*}{\tabincell{c} 37} \\
\midrule%\hline

\multirow{5}{*}{\tabincell{c} Software maintenance}
& Code clone detection  & 11 & \\ %\cline{2-13}
& Program repair & 6 & \\
& Code comment generation & 4 & \\
& Software quality evaluation & 4 & \\
& Source code representation & 4 & \multirow{-5}{*}{\tabincell{c} 28} \\
\midrule%\hline

\multirow{2}{*}{\tabincell{c} Software management}
%& Code in tree structure + text in token  & 4 & \\ %\cline{2-13}
& Software repository mining  & 19 & \\ %\cline{2-13}
& Effort cost prediction & 6 &  \multirow{-2}{*}{\tabincell{c} 25} \\
\bottomrule
\end{tabular}\vspace{-0.5cm}
\end{table*}

We summarized a set of research topics in which DL was engaged. Table~\ref{tab:table_topics} lists the research topics containing no less than three related studies. Software testing and debugging, as the most prevalent SE activity, has 37 primary studies in six topics. The most popular study is defect prediction (11 studies), followed by bug localization (7) and application testing (7). Software maintenance, as the second most popular activities, involves five research topics with 28 relevant studies, where code clone detection is the most popular research topic. Two SE activities, software implementation and software management, both contain two important research topics, where 19 primary studies mined software repositories by training DNNs, 6 studies estimated development cost, and 5 studies applied DL for code search. Software design and modeling only involve one popular topic, i.e., source code representation/modeling. There are no topics with more than three studies using DL techniques in software requirements.

%Another popular research topic in software maintenance is code analysis, containing several directions: code classification, code summarization, code smell detection, SATD detection, and code review.

\subsection{Software requirements}

%Requirements engineering encompasses a set of SE tasks that translate users' ideas and views towards software products into requirements  descriptions, aiming to achieve high-quality products and helping to specify the impact of the software on the organization, customers' needs, and how users will interact with the developed software.

\subsubsection{Requirements analysis}

A number of natural language-based documents that describe users' specific needs or services of a software product can be referred to as user requirements (aka, use cases, or actions) \cite{wiegers2013software}. %These user requirements are a representation of users' interactions with the product, mainly including four elements: use case name, actors denoting entities (i.e., persons or systems) that invokes the service to interact with the product, system boundary (i.e., system scope and boundary name), and relationship denoting four link types to other use cases (i.e., association, include, extend, and generalization).
Extracting use cases of a product from a large volume of textual requirement documentation is a common but labor-intensive task.  Since the manual mapping system states between requirements and simulation is a time-consuming task, Pudlitz et al. \cite{pudlitz2019extraction} proposed a self-trained Named-entity Recognition model combined with Bi-LSTM and CNN to extract the system states from requirements specification, working to reduce labor cost when linking the state extracted from requirements to the simulation signal.

%To alleviate this, Imam et al. \cite{al2018use} present a novel approach to automatically identify use cases in the requirements description of a software product. They first performed lexicons, syntaxes, and semantic analysis by using an NLP parser and then trained an ANN to extract actions (i.e., use cases) and actors on 5 datasets.

%\subsubsection{Requirement tracing}

%Wang et al. \cite{wang2018enhancing} trained a semantically enhanced  ANN to enhance the capability of automatically resolving polysemous terms by determining whether a term had the same meaning in different requirements. They validated the effectiveness of the proposed approach on two benchmark datasets and six long-lived open-source software projects. Their experimental results showed that compared with LSI, their model more accurately identified polysemous terms and increase the precision of requirements tracing. Experimental results based on 2,000 requirements documentation showed that their approach outperformed the state-of-the-art methods in terms of F1-score.

\subsubsection{requirement validation}

The requirements specification may be subject to validation and verification procedures, ensuring that developers have understood the requirements and the requirements conform to company standards. Winkler et al. \cite{winkler2019predicting} present an automatic approach to identify and determine the method for requirement validation. They predefined six possible verification methods and trained a CNN model as a multiclass and multilabel classifier to classify requirements with respect to their potential verification methods. The mixed results revealed that the imperfect training data impacted the performance of their classifier, but it still achieved good results on the testing data.

\subsection{Software design}

%Software design and modeling is a SE activity to express the software theory and algorithm and define software functions, objects, and the overall structure and interaction of source code so that the final functionality will satisfy user needs. In this section, we give a detailed description of each primary study used DL for software design and modeling.

\subsubsection{Software design patterns detection}

UI design is an essential component of software development, yet previous studies cannot reliably identify relevant high-fidelity UI designs from large-scale datasets. Martín et al. \cite{lopez2015neural} proposed a DL-based search engine to detect UI designs in various software products. The core idea of this search engine is to build a CNN-based wireframe image autoencoder to automatically generate labels on a large-scale dataset of Android UI designs. After manual evaluation of experimental results, they confirmed that their search engine achieved superior performance compared with image-similarity-based and component-matching-based methods. Thaller et al. \cite{thaller2019feature} proposed a flexible human- and machine-comprehensible software representation algorithm, namely Feature Maps. They first extracted subtrees from the system's abstract semantic graph (ASG). Then their algorithm pressed the high-dimensional and inhomogeneous vector space of these micro-structures into a feature map. Finally,  they adopted a classical machine learning model and a DL model (i.e., Random Forest and CNN) to identify instances of design patterns in source code. Their evaluation  suggested that Feature Map is an effective software representation method, revealing important information hidden in the source code.

\subsubsection{GUI modeling}

Chen et al. \cite{chen2018ui} proposed a neural machine translator to learn a crowd-scale knowledge of user interfaces (UI). Their generative tool encoded the spatial layouts of visual features learned from a UI image and learned to generate its graphical user interface (GUI) skeleton by combining RNN and CNN models. Its performance had been verified on the large-scale UI data from real-world applications. Moran et al. \cite{moran2018machine} proposed a strategy to facilitate developers automate the process of prototyping of GUIs in 3 steps: detection, classification, and assembly. First, they used computer vision techniques to detect logical components of a GUI from mock-up metadata. They then trained CNNs to category GUI-components into domain-specific types. Finally, a KNN algorithm was applied to generate a suitable hierarchical GUI structure to assemble prototype applications. Their evaluation  achieved an average GUI-component classification accuracy of 91\%.

\subsection{Software implementation}

%Software implementation is the process and act of writing source code, carrying out the instructions of development that allows computer software to function. This SE activity includes all SE issues included in the final manifestation of the software \cite{edward1995software} (e.g., code search, code generation, and code completion).

\subsubsection{Code search}

Gu et al. \cite{gu2016deep} proposed DeepAPI, a DL-based approach to generate functional API usage sequences for a given natural language-based user query by using an attention-based GRU Encoder-Decoder. DeepAPI first encoded the user query into a fixed-length context vector and produced the API sequence according to the context vector. It also enhanced their model by considering the importance of individual APIs. To evaluate its effectiveness, they empirically evaluated their approach on 7 million code snippets. Gu et al. \cite{gu2018deep} proposed a code search tool, DeepCS by using a novel DNN model. They considered code snippets as well as natural language descriptions, and then embedded them into a high-dimensional unified vector representation. Thus, DeepCS gave the relevant code snippets by retrieving the vector of the corresponding natural language query. They evaluated DeepCS with a large-scale dataset collected from GitHub.

Recently, several proposals use DL techniques for code search by embedding source code and given queries into vector space and calculating their semantic correlation \cite{bao2020psc2code}. Cambronero et al. \cite{cambronero2019deep} noticed that multiple approaches existed for searching related code snippets applied unsupervised techniques, while some adopted supervised ones to embed source code and queries for code search. They defined 3 RQs to investigate whether using supervised techniques is an effective way for code search and what types of DNNs and training corpus to use for this supervision. To understand these tradeoffs quantitatively, They selected and implemented four state-of-the-art code search techniques. They found that UNIF outperformed CODEnn and SCS models based on their benchmarks and suggested evaluating simple components first before integrating a complicated one.
Wan et al. \cite{wan2019multi} addressed the lack of analysis of structured features and inability to interpret search results in existing code search works. To address these issues, they presented a new approach MMAN, which adopted three different DNNs (i.e., LSTM, Tree-LSTM, and GGNN (Gated Graph Neural Network)) to analyze both shallow features and the semantic features in ASTs, and control-flow graphs (CFGs) of source code. The final results on a large-scale real-world dataset demonstrated that MMAN accurately provided code snippets. Huang et al. \cite{huang2020code} proposed an attention-based code-description representation learning model (CDRL) to refine the general DL-based code search approaches. They only picked up description terms and semantically related code tokens to embed a given query and its code snippet into a shared vector space. %CDRL performed a superior performance by 4-8\% in terms of precision, compared with the state-of-the-art approach.

\subsubsection{Programming}

%Some studies concentrated on proposing advanced DL-based models to provide convenience for developers for the sake of improving their programming efficiency, such as code generation, completion, localization, etc.

Gao et al. \cite{gao2019neural} introduced an attention-based Encoder-Decoder framework to directly generate sensible method names by considering the relationship between the functional descriptions and method names. To evaluate their model, experiments were performed on large-scale datasets for handling the cold-start problem, and the model achieved significant improvement over baselines.  Alahmadi et al. \cite{alahmadi2020code} applied a CNN model to automatically identify the exact location of code in images for reducing the noise. They extracted 450 screencasts covering C\#, Java, and Python programming languages to evaluate their model, and the final result showed that the accuracy of their model achieved 94\%. Wang et al. \cite{wang2019domain} proposed a Neural Network-based translation model to address the domain-specific rare word problem when carrying out software localization. They trained an RNN encoder-decoder framework and enhanced it by adding linguistic information. Nguyen et al. \cite{nguyen2018deep} proposed a DL-based language model, Dnn4C, which augmented the local context of lexical code elements with both syntactic and type contexts by using an FNN model. Empirical evaluation on code completion showed that Dnn4C improved accuracy by 24.9\%  on average over four baseline approaches.

\subsection{Software testing and debugging}

%Software testing and debugging is a process to evaluate the functionality of a software product with an intent to determine whether the product met the specified requirements or user needs. It also contributes to identifying the defects for ensuring the product is defect-free to implement the quality product. In this section, we will generalize specific primary studies into different SE tasks \cite{deshmukh2017towards, yan2018new} during the activity of testing and debugging.

\subsubsection{Defect prediction}

Defect prediction is the most extensive and active research topic in use of DL techniques in software maintenance. Almost 30\% of primary studies focused on identifying defects \cite{tong2018software, xu2019ldfr, wen2018well, liu2018connecting, barbez2019deep, dam2019lessons, zhou2019improving}.

\textbf{Metrics-based defect prediction.}
Metrics or features extracted from a software product can give a vivid description of its running state, and thus it is easy for researchers and participants to use these software metrics for defect prediction. Tong et al. \cite{tong2018software} proposed a novel two-stage approach, SDAEsT, which is build based on stacked denoising autoencoders (SDAEs) and a two-stage ensemble (TSE) learning strategy. Specifically, in the first stage, they used SDAEs to extract more robust and representative features. To mitigate the impact on the class imbalance problem and eliminate the overfitting problem, they propose TSE learning strategy as the second phase. They evaluated their work using 12 open-source defect datasets. %and verified the effectiveness of their proposed approach in terms of F1, AUC, and MCC. Xu et al.
Xu et al. \cite{xu2019ldfr} built an FNN model with a new hybrid loss function to learn the intrinsic structure and more discriminative features hidden behind the programs. Previous studies obtained process metrics throughout analyzing change histories manually and often ignored the sequence information of changes during software evaluation. For better utilization of such sequence data, Wen et al. \cite{wen2018well} built an RNN model to encode features from change sequences. They considered defect prediction as to the sequence labeling problem and performed fine-grained change analysis to extract six categories of change sequences, covering different aspects of software changes. Their evaluation on 10 public datasets showed that their approach achieved high performance in terms of F1-measure and AUC. To address the same problem, Liu et al. \cite{liu2018connecting} proposed to obtain the Historical Version Sequence of Metrics (HVSM) from various software versions as defect predictors and leveraged RNN to detect defects. Barbez et al. \cite{barbez2019deep} analyzed and mined the version control system to achieve historical values of structural code metrics. They then trained a CNN based classifier, CAME, to infer the anti-patterns in the software products. %Their experimental results indicated that the performance of CAME was superior to machine learning-based classifiers and state-of-the-art detection tools.

\textbf{Semantic-based defect prediction.}
%Since most of the existing studies detected defects in software products by manually designing features that can represent the characteristics of source code, encoding them into the fixed vector space, and training different learning algorithms. However, this traditional approach often lacked such a capability to capture and understand the semantic characteristics of programs for defect prediction. To bridge the gap between defect prediction features and programs' semantics,
Wang et al. \cite{wang2016automatically, wang2018deep} leveraged Deep Belief Network (DBN) to automatically learn semantic features from token vectors extracted from programs' ASTs, compared to most previous works that use manual feature specification. They evaluated their approach on file-level defect prediction tasks (within-project and cross-project) and change-level defect prediction tasks (within-project and cross-project) respectively. The evaluation results confirmed that DBN-based semantic features significantly outperformed the previous defect prediction based on traditional features in terms of F1-measure. Similarly, Dam et al. \cite{dam2019lessons} used a tree-based LSTM network, which can directly match with the AST of programs for capturing multiple levels of the semantics of source code.

\textbf{Just-In-Time (JIT) defect prediction.}
Hoang et al. \cite{hoang2019deepjit} presented an end-to-end DL-based framework, DeepJIT, for change-level defect prediction, or Just-In-Time (JIT) defect prediction. DeepJIT automatically extracted features from code changes and commit messages, and trained a CNN model to analyze them for defect prediction. The evaluation experiments on two popular projects showed that DeepJIT achieved improvements over 10\% for two open-source datasets in terms of AUC.

\subsubsection{Bug detection}

%Software aging caused by Aging-Related Bugs (ARBs) is prone to occur in the long-running system, leading to performance degradation and increasing failure rate during software execution. However, some files susceptible to ARB is small, limiting the availability of training data.
Wan et al. \cite{wan2019supervised} implemented a Supervised Representation Learning Approach (SRLA) based on an autoencoder with double encoding-layers to conduct cross-project Aging-Related Bugs (ARBs) prediction. They compared SRLA with the state-of-the-art approach, TLAP, to prove the effectiveness of SRLA. Wang et al. \cite{wang2019textout} present a novel framework, Textout, for detecting text-layout bugs in mobile apps. They formulated layout bug prediction as a classification issue and addressed this problem with image processing and deep learning techniques. Thus, they designed a specifically-tailored text detection method and trained a CNN classifier to identify text-layout bugs automatically. Textout achieved an AUC of 95.6\% on the dataset with 33,102 text-region images from real-world apps. Source code is composed of different terms and identifiers written in natural language with rich semantic information. Based on this intuition, Li et al. \cite{li2020deep} trained a DL-based model to detect suspicious return statements. They used a CNN to determine whether a given return statement in source code matched its method signature. To reduce the impact of the lack of negative training data, they converted the correct return statements in real-world projects to incorrect ones. Li et al. \cite{li2020predicting} proposed an AIOps solution for identifying node failures for an ultra-large-scale cloud computing platform at Alibaba.

\subsubsection{Vulnerability detection}

Dam et al. \cite{dam2018automatic} described a novel approach for vulnerability detection, which automatically captured both syntactic and semantic features of source code. The experiments on 18 Android applications and Firefox applications indicated that the effectiveness of their approach for within-project prediction and cross-project prediction. Tian et al. \cite{tian2020bvdetector} proposed to learn the fine-grained representation of binary programs and trained a Gated Recurrent Unit (BGRU) network model for intelligent vulnerability detection.  Han et al. \cite{han2017learning} trained a shallow CNN model to capture discriminative features of vulnerability description and exploit these features for predicting the multi-class severity level of software vulnerabilities. They collected large-scale data from the Common Vulnerabilities and Exposures (CVE) database to test their approach.

\subsubsection{Bug localization}

%Automatically localizing the potential buggy code snippets and files in software products can help and support practitioners to concentrate on critical files. Some primary studies used different DL techniques for bug localization \cite{lam2017bug}.
To locate buggy files, Lam et al. \cite{lam2017bug} built an autoencoder in combination with Information Retrieval (IR) technique, rVSM, which learned the relationship between the terms used in bug reports and code tokens in software projects. Some studies proposed to exploit CNN in the bug localization task \cite{zhang2019cnn, huo2019deep, xiao2019improving}. Zhang et al. \cite{zhang2019cnn} proposed CNNFL, which localized suspicious statements in source code responsible for failures based on CNN. They trained this model with test cases and tested it by evaluating the suspiciousness of statements. Huo et al. \cite{huo2019deep} present a deep transform learning algorithm, TRANP-CNN, for cross-project bug localization by training a CNN model to extract transferable semantic features from source code. Xiao et al. \cite{xiao2019improving} used the word-embedding technique to retain the semantic information of the bug report and source code and enhanced CNN to consider bug-fixing frequency and recency in company with feature detection techniques for bug localization. Li et al. \cite{li2019deepfl} proposed a novel approach, DeepFL, to learn latent features for precise fault localization, adopting an RNN model. The evaluation on the benchmark dataset, Defects4J, described that DeepFL significantly outperformed state-of-the-art approaches, i.e., TraPT/FLUCCS. Standard code parsers are of little help, typically resolving syntax errors and their precise location poorly. Santos et al. \cite{santos2018syntax} proposed a new methodology for locating syntax errors and provided some suggestions for possible changes for fixing these errors. Their methodology was of practical use to all developers but especially useful to novices frustrated with incomprehensible syntax errors.

\subsubsection{Test case generation}

%Some invalid text input of test cases may lead to detecting bugs in the product unsuccessfully, such as the E-mail address, website address, etc.
Liu et al. \cite{liu2017automatic} proposed a novel approach to automatically generate the most relevant text of test cases based on the context of use cases for mobile testing. Koo et al. \cite{koo2019pyse} implemented a novel approach, PySE1, to generate the test case. PySE1 tackled the limitations of symbolic execution schemes by proposing a DL-based reinforcement learning algorithm to improve the branch policy for exploring the worse case program execution. %The experimental results demonstrated that PySE handled the worst-case complexity of the program execution, such as maximum memory utilization.
Zhao et al. \cite{zhao2019seqfuzzer} trained a DL-based model that combines LSTM and FNN to learn the structures of protocol frames and deal with the temporal features of stateful protocols for carrying out security checks on industrial network and generating fake but plausible messages as test cases. Liu et al. \cite{liu2020deepsqli} proposed a deep natural language processing tool, DeepSQLi, to produce test cases used to detect SQLi vulnerabilities. They trained an encoder-decoder based seq2seq model to capture the semantic knowledge of SQLi attacks and used it to transform user inputs into new test cases.% Empirical results illustrated the effectiveness and the superiority of DeepSQLi compared with SQLmap.

\subsubsection{Program analysis}
%Program analysis refers, in general, to any examination of source code or program executions that attempt to find patterns or anomalies thought to reveal specific behaviors of the software. However, the performance of traditional program analysis techniques will be affected by different factors. Some primary studies thus enhanced traditional program analysis methods with DL techniques to address various SE tasks.
 % The complex setting is bound to reduce the precision of the scalable static analysis when identifying communication links between Android applications.
Program analysis refers to any examination of source code or program executions that attempt to find patterns or anomalies thought to reveal specific behaviors of the software.

\textbf{Static analysis.}
%applications can communicate with each other using an Android-specific message-passing system called Inter-Component Communication (icc). Misuse and abuse of icc may lead to several serious security vulnerabilities,
In Android mobile operating systems, applications communicate with each other a message-passing system, namely, Inter-Component Communication (ICC). Many serious security vulnerabilities may occur owing to misuse and abuse of communication links, i.e., ICCs. Zhao et al. \cite{zhao2018neural} presented a new approach to determine communication links between Android applications. They augmented static analysis with DL techniques by encoding data types of the links and calculating the probability of the link existence. To reduce the number of false alarms, Lee et al. \cite{lee2019classifying} trained a CNN model as an automated classifier to learn the lexical patterns in the parts of source code for detecting and classifying false alarms. %The results of their empirical evaluation suggested that their methodology was effective for detecting false positive alarms with 79.72\% of average precision.
Due to the impact of high false-positive rates on static analysis tools, Koc et al. \cite{koc2019empirical} performed a comparative empirical study of 4 learning techniques (i.e., hand-engineered features, a bag of words, RNNs, and GNNs) for classifying false positives, using multiple ground-truth program sets. Their results suggest that RNNs outperform the other studied techniques with high accuracy.

\textbf{Type inference.} %Type inference is limited in JavaScript since it was unable to handle duck-typing or runtime evaluation and generate richer compile-time information.
Helledoorn et al. \cite{hellendoorn2018deep} developed an automated framework, DeepTyper, a DL-based model to analyze JavaScript language and learn types that naturally occurred in certain contexts. It then provided a type of suggestion when the type checker cannot infer the types of code elements, such as variables and functions. Malik et al. \cite{malik2019nl2type} formulated the problem of inferring Javascript function types as a classification task. Thus, they trained a LSTM-based neural model to learn patterns and features from code annotated programs collected from real-world projects, and then predicted the function types of unannotated code by leveraging the learned knowledge.

\subsubsection{Testing techniques}

Many studies focus on new methods to perform testing, such as for apps \cite{pan2020reinforcement}, games \cite{zheng2019wuji}, and other software systems \cite{ben2016testing, chen2018drlgencert}. There are also some studies using well-known testing techniques (e.g., fuzzing \cite{godefroid2017learn, cummins2018compiler} and mutation testing \cite{mao2019extensive}) for improving the quality of software artifacts.
 %As is known to all that testing games is a challenging task due to a high dependence on manual playing and scripting based testing in the game industry. To fill this gap,
 Zheng et al. \cite{zheng2019wuji} conducted a comprehensive analysis of 1,349 real bugs and proposed Wuji, a game testing framework, which used an FNN model to perform automatic game testing. Ben et al. \cite{ben2016testing} also used the FNN to test Advanced Driver Assistance Systems (ADAS). They leveraged a multi-objective search to guide testing towards the most critical behaviors of ADAS. Pan et al. \cite{pan2020reinforcement} present Q-testing, a reinforcement learning-based approach, benefiting from both random and model-based approaches to automated testing of Android applications. Mao et al. \cite{mao2019extensive} performed an extensive study on the effectiveness and efficiency of the promising PMT technique. They also complemented the original PMT work by considering more features and the powerful deep learning models to speed up this process of generating the huge number of mutants.
 %Fuzzing is made up of repeatedly testing an application with modified, or fuzzed inputs to identify security vulnerabilities in input-parsing code.
 Godefroid et al. \cite{godefroid2017learn} used DL-based statistical machine-learning techniques to automatically generate input syntax suitable for input fuzzing.  Cummins et al. \cite{cummins2018compiler} introduced DeepSmith, a novel LSTM-based approach, for reducing the development task when using Fuzzers to discover bugs in compilers. They accelerated compiler validation through the inference of generative models for compiler inputs, and then applied DeepSmith to automatically generate tens of thousands of realistic programs. Finally, they constructed differential testing methodologies on these generated programs for exposing bugs in compilers. %Compared with the state-of-the-art approach, they noticed that DeepSmith significantly narrowed the time used for finding errors in compilers.

\subsubsection{Reverse execution}

A decompiler is a tool to reverse the compilation process for examining binaries. %This process is able to transform binaries into high-level code without the corresponding source code.
Lacomis et al. \cite{lacomis2019dire} introduced a novel probabilistic technique, namely Decompiled Identifier Renaming Engine (DIRE), which utilized both lexical and structural information recovered by the decompiler for variable name recovery. They also present a technique for generating corpora suitable for training and evaluating models of decompiled code renaming. Although reverse execution is an effective method to diagnose the root cause of software crashes, some inherent challenges may influence its performance. To address this issue, Mu et al. \cite{mu2019renn} present a novel DNN, which significantly increased the burden of doing hypothesis testing to track down non-alias relation in binary code and improved memory alias resolution. To achieve this, they first employed an RNN to learn the binary code pattern pertaining to memory access and then inferred the memory region accessed by memory references. Katz et al. \cite{katz2018using} noticed that the source code generated by decompilation techniques are difficult for developers to read and understand. To narrow the differences between human-written code and decompiled code, they trained a non-language-specific RNN model to learn properties and patterns in source code for decompiling binary code.

%\subsubsection{Others}

\subsection{Software maintenance}

%Software Maintenance is the process of modifying a software product after it has been compiled or delivered to the users. The majority purpose of software maintenance is to update and modify software applications after delivery to correct faults and to improve performance.
There are a lot of studies contributing to increasing maintenance efficiency, such as improving source code, logging information, software energy consumption, etc.  \cite{hoang2019patchnet, ma2019easy, liu2019variables, romansky2017deep, ha2019deepperf}. %In this section, we summarized a few research areas where DL techniques are often applied and give a brief description of each work.

\subsubsection{Code clone detection}

Code clone detection is a very popular SE task in software maintenance using DL, with around 20\% of primary studies concentrating on this research topic.

\textbf{RNN-based code clone detection.} Most studies use RNNs including RtNN \cite{gao2019teccd}, RvNN \cite{white2016deep}, and LSTM \cite{buch2019learning, perez2019cross, tufano2018deep} to identify clones in source code. White et al. \cite{white2016deep} proposed a novel code clone detector by combining two different RNNs, i.e., RtNN and RvNN, for automatically linking patterns mined at the lexical level with patterns mined at the syntactic level. They evaluated their DL-based approach based on file- and function-level. Gao et al. \cite{gao2019teccd} first transformed source code into AST by parsing programs and then adopted a skip-gram language model to generate vector representation of ASTs. After that, they used the standard RNN model to find code clones from java projects. Buch et al. \cite{buch2019learning} introduced a tree-based code clone detection approach, and traversed ASTs to form data sequences as the input of LSTM.  Perez et al. \cite{perez2019cross} also used LSTM to learn from ASTs, and then calculated the similarities between ASTs written in Java and Python for identifying cross-language clones. Since source code can be represented at different levels of abstraction: identifiers, Abstract Syntax Trees, Control Flow Graphs, and Bytecode, Tufano et al. \cite{tufano2018deep} conducted a series of experiments to demonstrate how DL can automatically learn code similarities from different representations.

\textbf{FNN-based code clone detection.} Some studies adopted FNNs for the code clone detection task \cite{li2017cclearner, nafi2019clcdsa, zhao2018deepsim}. Li et al. \cite{li2017cclearner} implemented a DL-based classifier, CClearner, for detecting function-level code clones by training an FNN. Compared with the approaches not using DL, CClearner achieved competitive clone detection effectiveness with a low time cost. Zhao et al. \cite{zhao2018deepsim} introduced a novel clone detection approach, which encoded data flow and control flow and into a semantic matrix and designed an FNN structure to measure the functional similarity between semantic representation of each code segment. Nafi et al. \cite{nafi2019clcdsa} proposed a cross-language clone detector without extensive processing of the source code and without the need to generate an intermediate representation. They trained an FNN model, which can learn different syntactic features of source code across programming languages and identified clones by comparing the similarity of features.

\textbf{Others.} % Apart from RNNs and FNNs, other DNNs were applied in clone detection \cite{yu2019neural, wang2020detecting, fang2020functional}.
For detecting Type-4 code clones, Yu et al. \cite{yu2019neural} present a new approach that uses tree-based convolution to detect semantic clones, by capturing both the structural information of a code fragment from its AST and lexical information from code tokens. They also addressed the limitation of an unlimited vocabulary of tokens and models. %The experimental results showed that their approach substantially outperformed an existing state-of-the-art approach with an increase of 0.42 and 0.15 in terms of F1-score on two popular code-clone benchmarks (OJClone and BigCloneBench).
Wang et al. \cite{wang2020detecting} developed a novel graph-based program representation method, flow-augmented abstract syntax tree (FA-AST), to better capture and leverage control and data flow information. FA-AST augmented original ASTs with explicit control and data flow edges and then adopted two different GNN models (i.e., gated graph neural network (GGNN) and graph matching network (GMN)) to measure the similarity of various code pairs. To effectively capture syntax and semantic information from programs to detect semantic clones, Fang et al. \cite{fang2020functional} adopted fusion embedding techniques to learn hidden syntactic and semantic features by building a novel joint code representation. They also proposed a new granularity for functional code clone detection called caller-callee method relationships. Finally, they trained a supervised deep learning model to find semantic clones.

\subsubsection{Code comment generation}

%Code comments, as one of the most important elements in source code, may determine the quality of software products. A project with good code comments benefits is conducive to programmers for update and evolution, and yet lacking code comments may cause new potential bugs in source code. To solve this problem, some studies proposed novel approaches by using DL techniques for generating code comments \cite{eberhart2020automatically, hu2020deep}.
Hu et al. \cite{hu2018deep} present a new approach that can automatically generate code comments for Java code to help developers better understand the functionality of code segments. They trained a LSTM-based framework to learn the program structure for better comments generation. %Experimental results demonstrate that their approach outperforms the state-of-the-art by a substantial margin.
The context information of the source code was not used and analyzed in previous automated comment summarization techniques. Ciurumelea et al. \cite{ciurumelea2020suggesting} proposed a semi-automated system to generate code comments by using LSTM. Zhou et al. \cite{zhou2019augmenting} combined program analysis and natural language processing to build a Dl-based seq2seq model to generate Java code comments.
To generate code summarization, Leclair et al. \cite{leclair2019neural} proposed a DL-based model combining texts from code with code structure from an AST. They processed data source as a separate input to reduce the entire dependence on internal documentation of code. %They observed improvement over two baseline techniques from SE literature and one from NLP literature.
Wan et al. \cite{wan2018improving} noticed that most of the previous work used the Encoder-Decoder architecture to generate code summaries, which omitted the tree structure of source code and introduced some bias when decoding code sequences. To solve these problems, they trained a deep reinforcement learning framework  that incorporated an abstract syntax tree structure as well as sequential content of code snippets. They trained this DNN by adopting an advantage reward composed of BLEU metric.

\subsubsection{Program repair}

Bhatia et al. \cite{bhatia2018neuro} proposed a novel neuro-symbolic approach combining DL techniques with constraint-based reasoning for automatically correcting programs with errors. Specifically, they trained an RNN model to perform syntax repairs for the buggy programs and ensured functional correctness by using constraint-based techniques.
Through evaluation, their approach was able to repair syntax errors in 60\% of submissions and identified functionally correct repairs for 24\% submissions.
Tufano et al. \cite{tufano2019empirical} proposed to leverage the proliferation of software development histories to fix common programming bugs. They used the Encoder-Decoder framework to translate buggy code into its fixed version after generating the abstract representation of buggy programs and their fixed code.
White et al. \cite{white2019sorting} trained an autoencoder framework to reason about the repair ingredients (i.e., the code reused to craft a patch). They prioritized and transformed suspicious statements and elements in the code repository for patch generation by calculating code similarities. Lutellier et al. \cite{lutellier2020coconut} present a new automated generate-and-validate program repair approach, CoCoNuT, which trained multiple models to extract hierarchical features and model source code at different granularity levels (e.g., statement and function level) and then constructed a CNN model to fix different program bugs. Liu et al. \cite{liu2019learning} proposed an automated approach for detecting and refactoring inconsistent method names by using Paragraph Vector and a CNN.
  Ni et al. \cite{ni2020analyzing} exploited the bug fixes of historical bugs to classify bugs into their cause categories based on the intuition that historical information may reflect the bug causes. They first defined the code-related bug classification criterion from the perspective of the cause of bugs and generated ASTs from diff source code to construct fixed trees. Then, they trained Tree-based Convolutional Neural Network (TBCNN) to represent each fixed tree and classified bugs into their cause categories according to the relationship between bug fixes and bug causes.

% \subsubsection{Code analysis}

% Source code analysis plays an important role in several maintenance tasks, which is used to analyze source code or compiled versions of code to help find security flaws. A set of specific SE tasks are involved, such as source code representation \cite{white2015toward, hussain2020codegru}, code classification \cite{dq2019bilateral, leclair2018adapting}, code summarization \cite{leclair2019neural, wan2018improving}, code smell detection \cite{fakhoury2018keep, liu2019deep}, SATD detection \cite{ren2019neural, zampetti2020automatically}, code review \cite{siow2020core, guo2019deep}, and impact analysis \cite{tufano2019learning}.

\subsubsection{Source code representation}

White et al. \cite{white2015toward} conducted an empirical study to adopt DL in software language modeling and highlight the fundamental differences between state-of-the-practice software language models and DL models. Their intuition is that the representation power of the abstractions is the key element of improving the quality of software language models. Therefore, the goal of this study was to improve the quality of the underlying abstractions by using Neural Network Language Models (i.e., Feed-forward neural networks (FNN), RNN) for numerous SE issues. They pinpointed that DL had a strong capability to model semantics and consider rich contexts, allowing it performed better at source code modeling. They evaluated these DL-based language models at a real SE task (i.e., code suggestion). Their evaluation results suggested that their model significantly outperformed the traditional language models on 16,221 Java projects.

 Hussain et al. \cite{hussain2020codegru} introduced a gated recurrent unit-based model, namely CodeGRU, to model source code by capturing its contextual, syntactical, and structural dependencies. The key innovation of their approach was to performing simple program analysis for capturing the source code context and further employed GRU to learn variable size context while modeling source code. They evaluated CodeGRU on several open-source java projects, and the experimental results verified that the approach alleviated the out of vocabulary issue. Using abstract syntax tree (AST)-based DNNs may induce a long-term dependency problem due to the large size of ASTs. To address this problem, Zhang et al. \cite{zhang2019novel} present an advanced AST-based Neural Network (ASTNN) for source code representation. The advanced ASTNN cut each entire AST into a set of small statement trees, and transform these subtrees into vectors by capturing the lexical and syntactical knowledge of statement trees. They applied a bidirectional RNN model to produce the vector representation of a code snippet. They used ASTNN to detect code clones and classify source code for evaluating its performance. Gill et al. \cite{gill2020thermosim} introduced a lightweight framework, ThermoSim, to simulate the thermal behavior of computing nodes in the Cloud Data Center (CDC) and measure the effect of temperatures on key performance parameters. They extended the previous framework, i.e., the CloudSim toolkit by presenting an RNN-based temperature predictor for helping to analyze the performance of some key parameters. The final results demonstrated that ThermoSim accurately modeled and simulated the thermal behavior of a CDC in terms of energy consumption, time, cost, and memory usage.

\subsubsection{Code classification.}

Bui et al. \cite{dq2019bilateral} described a framework of Bi-NN that built a neural network on top of two underlying sub-networks, each of which encoded syntax and semantics of code in a language.  Bi-NN was trained with bilateral programs that implement the same algorithms and/or data structures in different languages and then be applied to recognize algorithm classes across languages. Software categorization is the task of organizing software into groups that broadly describe the behavior of the software. However, previous studies suffered very large performance penalties when classifying source code and code comments only. Leclair et al. \cite{leclair2018adapting} proposed a set of adaptations to a state-of-the-art neural classification algorithm and conducted two evaluations. %The final result demonstrated the effectiveness of their proposed classifier.

%\subsubsection{Code summarization.}

\subsubsection{Code smell detection.}

Fakhoury et al. \cite{fakhoury2018keep} reported their experience in building an automatic linguistic anti-pattern detection using DNNs. They trained several traditional machine learning and DNNs to identify linguistic anti-patterns.  A big challenge for DL-based code smell detection is the lack of a large number of labeled datasets, and thus Liu et al. \cite{liu2019deep} present a DL-based approach to automatically generating labeled training data for DL-based classifiers. They applied their approach to detecting four common and well-known code smells, i.e., feature envy, long method, large class, and misplaced class.% The evaluation result suggested that the proposed approach significantly outperformed the state-of-the-art.

\subsubsection{Self-Admitted Technical Debt (SATD) detection.}

Technical debt (TD) is a metaphor to reflect the tradeoff developers make between short term benefits and long term stability. Self-admitted technical debt (SATD), a variant of TD, has been proposed to identify debt that is intentionally introduced during SDLC. Ren et al. \cite{ren2019neural} proposed a CNN-based approach to determine code comments as SATD or non-SATD. They exploited the computational structure of CNNs to identify key phrases and patterns in code comments that are most relevant to SATD for improving the explainability of our model’s prediction results. Zampetti et al. \cite{zampetti2020automatically} proposed to automatically recommend SATD removal strategies by building a multi-level classifier on a curated dataset of SATD removal patterns. Their strategy was capable of recommending six SATD removal patterns, i.e., changing API calls, conditionals, method signatures, exception handling, return statements, or telling that a more complex change is needed. %The experimental result revealed that SATD removal followed recurrent patterns and indicated the feasibility of supporting developers in this task with automated recommenders.

\subsubsection{Code review.}

Siow et al. \cite{siow2020core} believed that the hinge of the accurate code review suggestion is to learn good representations for both code changes and reviews. Therefore, they designed a multi-level embedding framework to represent the semantics provided by code changes and reviews and then well trained through an attention-based deep learning model, CORE. Guo et al. \cite{guo2019deep} proposed Deep Review Sharing, a new technique based on code clone detection for accurate review sharing among similar software projects, and optimized their technique by a series of operations such as heuristic filtering and review deduplication. They evaluated Deep Review Sharing on hundreds of real code segments and it won considerable positive approvals by experts, illustrating its effectiveness.

\subsubsection{Software quality evaluation}

Variety evaluation metrics can be used to describe the quality of software products \cite{rani2018neural}.

%\textbf{Traceability.} Guo et al. \cite{guo2017semantically} present a solution to incorporate requirements artifact semantics and domain knowledge into the tracing solution for avoiding misunderstanding of software artifacts and delivering imprecise and inaccurate results. They implemented a tracing neural network architecture to generate trace links by using word Embedding and various RNN models. They evaluated their approach on 360 different configurations of the tracing network and found that BI-GRU was the best DNN model for the tracing task.

\textbf{Software trustworthiness.} It is essential and necessary to evaluate software trustworthiness based on the influence degrees of different software behaviors for minimizing the interference of human factors. Tian et al. \cite{tian2020software} constructed behaviour trajectory matrices to represent the behaviour trajectory and then trained the deep residual network (ResNet) as a software trustworthiness evaluation model to classify the current software behaviors. After that, they used the cosine similarity algorithm to calculate the deviation degree of the software behavior trajectory.

\textbf{Readability.} Mi et al. \cite{mi2018improving} proposed to leverage CNN to improve code readability classification. First, they present a transformation strategy to generate integer matrices as the input of ConvNets. Then they trained Deep CRM, a DL-based model, which was made up of three separate ConvNets with identical architectures for code readability classification.

\textbf{Maintainability.} Kumar et al. \cite{kumar2016hybrid} performed two case studies and applied three DNNs i.e., FLANN-Genetic (FGA and AFGA), FLANN-PSO (FPSO and MFPSO), FLANN-CSA (FCSA), to design a model for predicting maintainability. They also evaluated the effectiveness of feature reduction techniques for predicting maintainability. The experimental result showed that feature reduction techniques can achieve better results compared with using DNNs. %\textbf{Reliability} Rani et al.  proposed a learning  algorithm of supervised back‐propagation neural networks

% \subsubsection{Bug report summarization}

% Li et al. \cite{li2018unsupervised} performed the first exploration on bug report summarization by applying DL techniques. They proposed a new DL-based framework, called Deepsum, which used a  stepped auto-encoder to integrate the features of bug reports into the auto-encoder network. This framework was an unsupervised DNN, reducing the effort on labeling datasets.

\subsection{Software management}

%Software management involved a series of SE tasks dedicating to time management, planning, scheduling, resource allocation, etc.. Besides, it also involved some issues that mined useful knowledge from software repositories for fully understanding the user and developer behaviors. In this section, we will mainly introduce two main research topics related to software management.

\subsubsection{Effort estimation}

Since only 39\% of software projects are finished and published on time relative to the duration planned originally \cite{lopez2015neural, bisi2016software}, it is necessary to assess the development cost to achieve reliable software within development schedule and budget. Lopez et al. \cite{lopez2015neural} compared three different neural network models for effort estimation. The experimental result demonstrated that MLP and RBFNN can achieve higher accuracy than the MLR model. Choetkiertiku et al. \cite{choetkiertikul2018deep} observed that few studies focused on estimating effort cost in agile projects, and thus they proposed a DL-based model for predicting development cost based on combining two powerful DL techniques: LSTM and recurrent highway network (RHN). %They performed comprehensive experiments on large-scale datasets to demonstrate its effectiveness.
Phannachitta et al. \cite{phannachitta2020optimal} conducted an empirical study to revisit the systematic comparison of heuristics-based and learning-based software effort estimators on 13 standard benchmark datasets. Ochodek et al.\cite{ochodek2020deep} employed several DNNs (i.e., CNN, RNN, Convolutional + Recurrent Neural Network (CRNN)) to design a novel prediction model, and compared the performance of the DL-based model with three state-of-the-art approaches: AUC, AUCG, and BN-UCGAIN. They noticed that CNN obtained the best prediction accuracy among all software effort adaptors.

\subsubsection{Software repository mining}

Some primary studies use DL techniques to mine the contents in different software repositories \cite{le2018deep, mahadi2020cross, guo2019systematic}. In this section, we introduce three most widely used repositories, i.e., Stack Overflow (SO) \cite{liu2018fasttagrec, zhou2019deep}, GitHub \cite{jiang2017automatically}, and Youtube \cite{ott2018deep}.

\textbf{Mining Stack Overflow (SO).} The questions and answers in SO contain a great deal of useful information that is beneficial for programmers to address some tough problems when implementing software products. Considering a question and its answers in Stack Overflow as a knowledge unit, Xu et al. \cite{xu2016predicting, xu2018prediction} extracted the knowledge units and analyzed the potential semantic relationship between Q and A in each unit. They formulated the problem of predicting semantically linkable knowledge units as a multi-class classification problem and adopted a CNN model combining with word-embedding to capture and classify document-level semantics of knowledge units. Chen et al. \cite{chen2016learning} also applied word embeddings and the CNN model to mine SO for retrieving cross-lingual questions. They compared their approach with other translation-based methods, and the final results showed that their approach can significantly outperform baselines and can also be extended to dual-language document retrieval from different sources.  Yin et al. \cite{yin2018learning} proposed a new approach to pair the title of a question with the code in the accepted answer by mining high-quality aligned data from SO using two sets of features. To validate whether DL was the best solution in all research topics, Menzies et al. \cite{menzies2018500+} conducted a case study to explore faster approaches for text mining SO. They compared nine different solutions including traditional machine learning algorithms and DL algorithms and noticed that a tuned SVM performs similarly to a deep learner and was over 500 times faster than DL-based models. %To conclude the main challenges when training a DL model,
Zhang et al. \cite{zhang2019empirical} performed an empirical study to mine the posts in SO for investigating what potential challenges developers face when using DL. They also built a classification model to quantify the distribution of different Sort of DL related questions. They summarized the three most frequently asked questions and provided five future research directions. Since answers to a question may contain extensive irrelevant information, Wang et al. \cite{wang2019extracting} proposed a DL-based approach, DeepTip, using different CNN architectures to extract short practical and useful tips and filtered useless information from developer answers. They conducted a user study to prove the effectiveness of their approach and the extensive empirical experiments demonstrated that DeepTip can extract useful information from answers with high precision and coverage, significantly outperforming two state-of-the-art approaches.% by up to 56.7\% and 162\% in precision.

\textbf{Mining GitHub.} Huang et al. \cite{huang2018automating} proposed a new model to classify sentences from issue reports of four projects in GitHub. They constructed a dataset collecting 5,408 sentences and refined previous categories (i.e., feature request, problem discovery, information seeking, information giving, and others). They then trained a CNN-based model to automatically classify sentences into different categories of intentions and used the batch normalization and automatic hyperparameter tuning approach to optimize their model. %Compared with baseline, their model improved the accuracy by 171\%.
Xie et al. \cite{xie2019deeplink} proposed a new approach to recover issue-commit links. They constructed the code knowledge graph of a code repository and captured the semantics of issue- or commit-related text by generating embeddings of source code files. Then they trained a DNN model to calculate code similarity and semantic similarity using additional features. Ruan et al.\cite{ruan2019deeplink} propose a novel semantically-enhanced link recovery method, DeepLink, using DL techniques.
They applied word embedding and RNN to implement a DNN architecture to learn the semantic representation of code and texts as well as the semantic correlation between issues and commits. They compared DeepLink with the state-of-the-art approach on 10 GitHub Java projects to evaluate the effectiveness of DeepLink. Liu et al. \cite{liu2019automatic}  proposed a DL-based approach to automatically generate pull request descriptions based on the commit messages and the added source code comments in pull requests. They formulated this problem as a text summarization problem and solved it, constructing an attentional encoder-decoder model with a pointer generator.

\textbf{Mining Youtube.} Ott et al. \cite{ott2018deep} employed CNN and AutoEncoder to identify Java code from videos on Youtube. Their approach was able to identify the presence of typeset and handwritten source code in thousands of video images with 85.6\%-98.6\% accuracy based on syntactic and contextual features learned through DL techniques. Zhao et al. \cite{zhao2019actionnet} present a new technique for recognizing workflow actions in programming screencasts. They collected programming screencasts from Youtube and trained a CNN model to identify nine classes of frequent developer actions by employing the image differencing technique and training a CNN model.

\begin{tcolorbox}[breakable,colback=gray!20,colframe=gray!35!black,title=Summary]
\small
 \begin{enumerate}
     \item We grouped six SE activities based on the body knowledge of SE -- Software requirements, Software design and modeling, Software implementation, Software testing and debugging, Software maintenance, and Software management -- and provided an outline of the application trend of DL techniques among these SE activities.
     \item We summarized various SE tasks into four categories -- regression task, classification task, ranking task, and generation task -- and classified all primary studies based on the task types. Most studies can be mapped to classification tasks, and only 11\% of primary studies are mapped to ranking tasks.
     \item Software testing and software maintenance are the two SE activities containing the most related studies and include 17 specific SE research topics in which DL techniques were used.
    % \item We classified each primary study task into suitable topics and gave a brief description of each SE task to help researchers understand the application status of DL in different topics.
 \end{enumerate}
\end{tcolorbox}

\section{Limitations}

\vspace{0.1cm}\noindent\textbf{Data Extraction.} There are several potential limitations to our work. One limitation is the potential bias in data collection. Although we have listed the data items used for analysis in RQs in Section 3.4, some disagreements still appeared inevitably when extracting related content and classifying these data items. Two researchers first recorded the relevant descriptions in each primary study and then discussed and generalized temporary categories based on all data in one item by comparing the objectives and contributions with related studies. If they were unable to reach an agreement on classification, another researcher would join in and resolve differences, which can mitigate the bias in data extraction to study validity.

\vspace{0.1cm}\noindent\textbf{Study Selection Bias.} Another threat might be the possibility of missing DL related studies during the literature search and selection phase. We are unable to identify and retrieve all relevant publications considering the many publication venues publishing SE relevant papers. Therefore, we carefully selected 21 publication venues, including conference proceedings, symposiums, and journals, which can cover many prior works in SE. Besides, we identified our search terms and formed a search string by combining and amending the search strings used in other literature reviews on DL. These could keep the number of missed primary studies as small as possible. 

\section{Challenges and opportunities}

%\subsection{Challenges}

%\subsubsection{The generalization of DL.}

\vspace{0.1cm}\noindent{\bf Using DL in more SE activities. }
We found that DL has been widely used in certain SE topics, such as defect prediction, code clone detection, software repository mining, etc. However, few studies used DL for some SE research topics compared with other techniques or other learning algorithms. Although Software requirements and software design are the most two important documentations during SDLC, not many studies focus on these two SE activities. Therefore, one potential opportunity is that researchers can utilize DL techniques to explore new research topics or pay attention to classical topics in software requirements and design.

\vspace{0.1cm}\noindent{\bf The transparency of DL. }
In this study, we discussed 142 studies that used DL to address various SE issues. We noticed that few studies declared the reason for the architecture they chose and explained the necessity and value of each layer in DNN, which leads to low transparency of the proposed DL solutions. Because it is inherently difficult to comprehend what drives the decisions of the DNN due to the black-box nature of DL. Humans only pay attention to the output of DNNs since they can provide wise and actionable suggestions for humans. Furthermore, DL algorithms sift through millions of data points to find patterns and correlations that often go unnoticed by human experts. The decision they make based on these findings often confound even the engineers who created them. New methods and studies on explaining the decision-making process of DNNs should be an active research direction, which facilitates software engineers to design more effective DNN architectures for specific SE problems.

% \subsection{The replicability and reproducibility of DL.}

% Replicability and reproducibility are widely accepted as key considerations in scientific research. However, different from ML algorithms, the strong randomness in model initialization and optimization makes most DL-based models are difficult to exact reproduction, and thus many studies often do not perform experiments to evaluate these two factors. Besides, it is difficult to replicate and reproduce DL algorithms by manually setting model parameters exactly. Some studies realized the importance of these two issues, but they considered them as mere threats and left them in future work. Recently, only a few works investigated the replicability and reproducibility of DL in SE. Liu et al. \cite{liu2020replicability} conducted a SLR on 93 studies to investigate these two issues. They observed that over 74.2\% of the studies did not share any source code and datasets to support the replicability of DL models. They also noticed that a set of factors were likely to result in these two factors significantly influenced, e.g., unstable optimization process, non-convergent models, the small size of testing data, etc. New algorithms are needed to enhance DL-based solution stability and convergence and avoid performance sensitivity on different sampled data.

% \subsection{DL algorithm selection.}

% In RQ2, we summarized three types of algorithm selection strategies that were maily adopted by most of primary studies, and also briefly analyzed the usage of these strategies. However,

\vspace{0.1cm}\noindent{\bf DL in real scenarios. }
We analyzed the source of datasets used for training DNNs in RQ3 and noticed that only 4\% studies used industry datasets to train and evaluate their proposed models. In fact, most studies contribute to addressing real-world SE issues, but the novel solutions or DL-based models have not been evaluated on industry data. There is a room for more industry-academia collaboration so that the proposed models can be validated on real industry data (which can be of very different nature than open-source data.

\vspace{0.1cm}\noindent{\bf Data hungry.} 
When analyzing the studies related to code clone detection, we found that several open-source public data sets are often used repeatedly in these studies to evaluate their proposed models. A similar situation also exists in other research topics. These highlight the dependence on some studies on large publicly available labelled datasets. One reason is that training a DNN requires a massive volume of data, but copious amounts of training data are rarely available in most SE tasks. Besides, it is impossible to give every possible labeled sample of a problem space to a DL algorithm. Therefore, it will have to generalize or interpolate between its previous samples in order to classify data it has never seen before. It is a challenge to tackle the problem that DL techniques currently lack a mechanism for learning abstractions through explicit, verbal definition and only work best with thousands, millions, or even billions of training examples. One solution is to construct widely accepted datasets by using industrial labeled data or crawling software repositories to collect related data samples and label them as public datasets. Another is to develop new DL techniques, which can learn how to learn and be trained as an effective model with as little data size as possible, such as meta-learning.

% \subsection{Stability of DL}

% DL-based models are highly susceptible to noise. Small variations in the input data can lead to drastically different results, making them inherently unstable. unstable DL-based models may arrive at completely incorrect insights and even may introduce errors or defeats in software systems when processing certain important SE tasks.

%\subsection{Opportunities}

% \subsection{SE for DL.}

% There are many challenges with building production-ready systems with DL components, especially if the company does not have a large research group and a highly developed supporting infrastructure. Therefore, guaranteeing the quality of these components has become a key step after building a DL system. Nowadays, although a few studies detected the defects in DL components with several software testing techniques, the work of the quality assessment and guarantee towards DL systems is still incomplete. based on this observation, another potential research topic is to utilize various testing and maintenance techniques in SE for evaluating and enhancing existing DL systems.

\vspace{0.1cm}\noindent{\bf Performance of DL and traditional techniques.}
DL has been gradually used in more and more SE tasks, replacing the status of traditional algorithms. However, are DL algorithms really more efficient than traditional algorithms? What SE tasks are suitable for DL algorithms? What factors determine whether DL algorithms are better or worse than traditional algorithms? These questions are almost unanswered and neglected by most researchers. A potential opportunity is to answer these questions. researchers can conduct empirical studies to investigate what SE tasks or environments are suitable for DL and compare the performance between DL and traditional techniques in some important SE research topics where most of Dl algorithms were applied.

%deep learning 真的比 ML , traditional methods 好么

\section{Conclusion}
This work performed a SLR on 142 primary studies related to DL for SE from 21 publication venues, including conference proceedings, symposiums, and journals. We established four research questions to comprehensively investigate various aspects pertaining to applications of DL models to SE tasks. Our SLR showed that there was a rapid growth of research interest in the use of DL for SE.  Through an elaborated investigation and analysis, three DL architectures containing 30 different DNNs were used in primary studies, where RNN, CNN, and FNN are the three most widely used neural networks compared with other DNNs. We also generalized three different model selection strategies and analyzed the popularity of each one. To comprehensively understand the DNN training and testing process, we provided a detailed overview of key techniques in terms of data collection, data processing, model optimization, and model evaluation in RQ3. In RQ4, we analyzed the distribution of DL techniques used in different SE activities, classified primary studies according to specific SE tasks they solved and gave a brief summary of each work. We observed that DL techniques were applied in 23 SE topics, covering 6 SE activities. Finally, we identified a set of current challenges that still need to be addressed in future work on using DLs in SE.

%Besides, we noticed that most of the novel DL-based models were published in conference papers by analyzing the publication trend in the last five years.

%
% The next two lines define the bibliography style to be used, and the bibliography file.
\bibliographystyle{ACM-Reference-Format}
\bibliography{sample-acmsmall}

%
% If your work has an appendix, this is the place to put it.
% \appendix

% \section{Research Methods}

% \subsection{Part One}

% Lorem ipsum dolor sit amet, consectetur adipiscing elit. Morbi malesuada, quam in pulvinar varius, metus nunc fermentum urna, id sollicitudin purus odio sit amet enim. Aliquam ullamcorper eu ipsum vel mollis. Curabitur quis dictum nisl. Phasellus vel semper risus, et lacinia dolor. Integer ultricies commodo sem nec semper.

% \subsection{Part Two}

% Etiam commodo feugiat nisl pulvinar pellentesque. Etiam auctor sodales ligula, non varius nibh pulvinar semper. Suspendisse nec lectus non ipsum convallis congue hendrerit vitae sapien. Donec at laoreet eros. Vivamus non purus placerat, scelerisque diam eu, cursus ante. Etiam aliquam tortor auctor efficitur mattis.

% \section{Online Resources}

% Nam id fermentum dui. Suspendisse sagittis tortor a nulla mollis, in pulvinar ex pretium. Sed interdum orci quis metus euismod, et sagittis enim maximus. Vestibulum gravida massa ut felis suscipit congue. Quisque mattis elit a risus ultrices commodo venenatis eget dui. Etiam sagittis eleifend elementum.

% Nam interdum magna at lectus dignissim, ac dignissim lorem rhoncus. Maecenas eu arcu ac neque placerat aliquam. Nunc pulvinar massa et mattis lacinia.

\end{document}